\documentclass[lettersize,journal]{IEEEtran}

\usepackage{times}
\usepackage{amsmath,dsfont}
\usepackage{amssymb,amsthm}
\usepackage{epsfig,verbatim}
\usepackage{subfigure}
\usepackage{setspace}
\usepackage{color}
\usepackage{cite}
\usepackage{epstopdf}
\usepackage{graphics}
\usepackage{accents}
\usepackage{acronym}
\usepackage[bookmarks,colorlinks]{hyperref}
\usepackage{booktabs}
\usepackage{mathtools}
\usepackage{float} 
\usepackage{algpseudocode}
\usepackage{enumitem}
\makeatletter
\newcommand{\removelatexerror}{\let\@latex@error\@gobble}
\makeatother

\def\BibTeX{{\rm B\kern-.05em{\sc i\kern-.025em b}\kern-.08em
    T\kern-.1667em\lower.7ex\hbox{E}\kern-.125emX}}
 \usepackage{setspace}
\makeatletter
\newcommand{\oset}[3][0ex]{%
  \mathrel{\mathop{#3}\limits^{
    \vbox to#1{\kern-2\ex@
    \hbox{$\scriptstyle#2$}\vss}}}}
\makeatother

\usepackage[linesnumbered,ruled,vlined]{algorithm2e}
\SetKwInput{KwData}{\textbf{Init}} 
\let\oldnl\nl
\newcommand{\nonl}{\renewcommand{\nl}{\let\nl\oldnl}}
\newcommand{\myVec}[1]{{\boldsymbol{#1}}}

\newcommand{\mySet}[1]{\mathcal{#1}}
	
\newcommand{\mys}{{\myVec{s}}}	

\newcommand{\Weights}{\myVec{\varphi}}
\newcommand{\TrainSet}{\mySet{M}^{\rm tr}}

\newcommand{\Blklen}{B}			 			

\makeatother

\newtheorem{corollary}{Corollary}
\newtheorem{proposition}{Proposition}

\acrodef{fc}[FC]{fully-connected}
\acrodef{adc}[ADC]{analog-to-digital convertor}
\acrodef{cs}[CS]{compressed sensing}
\acrodef{dtft}[DTFT]{discrete-time Fourier transform}
\acrodef{dnn}[DNN]{deep neural network} 
\acrodef{csi}[CSI]{channel state information}
\acrodef{bpsk}[BPSK]{binary phase shift keying}
\acrodef{map}[MAP]{maximum a-posteriori probability}
\acrodef{snr}[SNR]{signal-to-noise ratio}
\acrodef{bs}[BS]{base station} 
\acrodef{iot}[IOT]{Interent of Things}
\acrodef{mimo}[MIMO]{multiple-input multiple-output}
\acrodef{siso}[SISO]{single-input single-output}
\acrodef{mse}[MSE]{mean-squared error}
\acrodef{pdf}[PDF]{probability density function}
\acrodef{rv}[RV]{random variable}
\acrodef{ml}[ML]{machine learning}
\acrodef{mc}[MC]{monte carlo}
\acrodef{fec}[FEC]{forward error correction}
\acrodef{qpsk}[QPSK]{quadrature phase shift keying}
\acrodef{rs}[RS]{Reed-Solomon}
\acrodef{lti}[LTI]{linear time-invariant}
\acrodef{wss}[WSS]{wide-sense stationary}
\acrodef{psd}[PSD]{power spectral density}
\acrodef{ser}[SER]{symbol error rate} 
\acrodef{ber}[BER]{bit error rate} 
\acrodef{gd}[GD]{gradient descent}
\acrodef{sgd}[SGD]{stochastic gradient descent} 
\acrodef{isi}[ISI]{intersymbol interference}  
\acrodef{awgn}[AWGN]{additive zero-mean white real Gaussian noise} 
\acrodef{ut}[UT]{user terminal} 
\acrodef{mmw}[mmWave]{millimeter wave}
\acrodef{noma}[NOMA]{non-orthogonal multiple access}
\acrodef{mac}[MAC]{mulitple access channel}
\acrodef{fl}[FL]{Federated learning}
\acrodef{lstm}[LSTM]{long short-term memory}
\acrodef{maml}[MAML]{model-agnostic meta-learning}
\acrodef{sic}[SIC]{soft interference cancellation}
\acrodef{pmf}[PMF]{probability mass function}
\acrodef{crc}[CRC]{cyclic redundancy check}
\acrodef{ece}[ECE]{expected calibration error}
\acrodef{lbd}[LBD]{learnable Bernoulli dropout}
\acrodef{arm}[ARM]{Augment-REINFORCE-Merge}
\acrodef{kl}[KL]{Kullback-Leibler}
\acrodef{ai}[AI]{artificial intelligence}
\acrodef{ddm}[DDM]{Drift Detection Method}
\acrodef{pht}[PHT]{Page Hinkley Test}
\acrodef{rnn}[RNN]{recurrent neural network}
\acrodef{cnn}[CNN]{convolutional neural network}

\IEEEoverridecommandlockouts 

\title{Asynchronous Online Adaptation via Modular Drift Detection for Deep Receivers}
\author{  
	\IEEEauthorblockN{Nicole Uzlaner$^*$, Tomer Raviv$^*$,  Nir Shlezinger, and Koby Todros
	} 
	\thanks{
		Parts of this work were accepted for presentation in the 2024 IEEE Signal Processing Advances in Wireless Communications (SPAWC) as the paper~\cite{uzlaner2024concept}. This work was supported by the Israeli Innovation Authority. 
 The authors are with the School of ECE, Ben-Gurion University of the Negev, Israel (e-mail: \{nicoleu; ravivto\}@post.bgu.ac.il, \{nirshl; todros\}@bgu.ac.il).}
}
\begin{document}
    \maketitle
    \def\thefootnote{*}\footnotetext{These authors contributed equally to this work}
	\begin{abstract} 
Deep learning is envisioned to facilitate the operation of wireless receivers, with emerging architectures integrating \acp{dnn} with traditional modular receiver processing.  While deep receivers were shown to operate reliably in complex settings for which they were trained, the dynamic nature of wireless communications gives rise to the need to repeatedly adapt deep receivers to channel variations. However, frequent re-training is costly and ineffective, while in practice, not every channel variation necessitates adaptation of the entire \ac{dnn}. 
In this paper, we study {\em concept drift detection}  for identifying when does a deep receiver no longer match the channel, enabling {\em asynchronous adaptation}, i.e., re-training only when necessary. We identify existing drift detection schemes from the machine learning literature that can be adapted for deep receivers in dynamic channels, and propose a novel soft-output detection mechanism tailored to the communication domain. 
Moreover, for deep receivers that preserve conventional modular receiver processing, we design modular drift detection mechanisms, that simultaneously identify {\em when} and {\em which} sub-module to re-train. 
The provided numerical studies show that even in a rapidly time-varying scenarios, asynchronous adaptation via modular drift detection dramatically reduces the number of trained parameters and re-training times, with little compromise on performance.
\end{abstract}


 \acresetall
\section{Introduction}

The constantly growing demands for throughput and robustness led to a rising interest in designing \ac{dnn} for wireless communications~\cite{saad2019vision}.
 A key aspect in which \acp{dnn} are expected to contribute is receiver processing, where their abstractness can be leveraged to learn from data to reliably operate in complex and unknown channel models~\cite{dai2020deep,mao2018deep,tong2022nine}. 
Still, integrating \acp{dnn} into receiver processing gives rise to several core challenges associated with such systems, that are notably more dominant in wireless communications compared to traditional deep learning domains, e.g., computer vision and  language processing. These fundamental challenges include the {\em dynamic nature} of wireless channels, and the {\em limited compute/power resources} of wireless devices~\cite{raviv2023adaptive}.

To see how these challenges affect deep receivers, recall that \acp{dnn} are typically trained offline on powerful servers for a task determined by the training data and its distribution. In wireless communications, the dynamic nature of the channel implies that the receiver task (which is dictated by the data distribution)  changes over time~\cite{aoudia2021end}. The direct approach to tackle this, coined {\em joint learning}~\cite{oshea2017introduction}, trains over a broad range of channel conditions. Joint learning effectively seeks a non-coherent receiver, at the cost of performance degradation compared with coherent operation. 
A second approach estimates the channel on each coherence interval, providing the estimate as an additional input to the \ac{dnn}~\cite{honkala2021deeprx, goutay2021machine}, while being typically limited 
to linear channel models. 
A third approach designs deep receivers that can be applied to several different data distributions observed during training using architectures based on ensemble learning~\cite{raviv2020data} and hypernetworks~\cite{goutay2020deep,liu2024hypernetwork}. 

The aforementioned approaches train \ac{dnn}-aided receivers as in conventional deep learning domains, i.e., offline and not on-device. 
An alternative approach, which is promising in terms of performance and flexibility,  re-trains the \ac{dnn} on device online. This leads to improved performance as the deep receiver matches the current data distribution, particularly when re-training synchronously with channel variations, e.g., on each coherence duration~\cite{shlezinger2019viterbinet,shlezinger2019deepsic}. Yet, online learning is challenging to implement due to the limited resources of wireless receives. Various techniques have been proposed to facilitate online learning by designing rapid training mechanisms~\cite{raviv2022online, park2020learning}, combining Bayesian training to facilitate learning from scarce pilots~\cite{zecchin2023robust, cohen2022bayesian,raviv2023modular}, and developing data generation and enrichment techniques~\cite{raviv2022data, fischer2022adaptive}.  
Alternative approaches design deep receivers by converting conventional modular receiver processing into \ac{ml} models~\cite{shlezinger2020model,shlezinger2021model}, obtaining modular architectures that are more compact compared to end-to-end \acp{dnn} and are  more amenable to rapid adaptation with limited data~\cite{farsad2020data,zappone2019wireless,raviv2023adaptive}.

The approaches discussed above can still be computationally infeasible if applied repeatedly on each coherence interval. Moreover, as long as the the distribution on which the \ac{dnn}-aided receiver was trained matches the instantaneous test distribution, the receiver is fit to successfully detect and re-training is not mandatory. For example, in the studies ~\cite{shlezinger2019viterbinet,shlezinger2019deepsic,farsad2018neural}, it was empirically observed that there are cases in which mismatches between the train channel and test channel lead to performance degradation, while in other cases \ac{dnn}-aided receivers generalize well to unseen channels. 

Taking the above into consideration, we claim that the rate of adaptations can be reduced if done {\em asynchronously}, training only when the change in the channel causes performance deterioration. Moreover, the emergence of modular deep receiver architectures implies that one can possibly adapt only some of the sub-modules of the overall architecture,   reducing the computational effort of each online adaptation. This motivates exploring ways to enable asynchronous and modular online learning by identifying when re-training is needed and which sub-module should be adapted, in order to enable  high performance  online learning without inducing notable excessive complexity. 
 
In this work, we study mechanisms for detecting when and which sub-modules of a deep receiver should be re-trained under dynamic channel conditions. We seek to  reduce the amount of re-training, as well as the number of parameters adapted, 
while keeping the overall \ac{ser} similar to  frequent online training.  We adopt the framework of {\em concept drift}, which encompasses \ac{ml} techniques for detecting changes in the suitability of a model due to variations in data distributions~\cite{lu2018learning}. The fact that concept drift detectors typically focus on identifying {\em shifts in the usefulness of an \ac{ml} model}, rather than identifying {\em shifts in the underlying distribution}~\cite{bayram2022concept}, makes this framework suitable for realizing asynchronous online learning. We thus seek to enable deep receivers to operate reliably while avoiding re-training the overall \ac{dnn} on every coherence duration.

We commence by exploring techniques for identifying when re-training is needed, without accounting for the architecture of the deep receiver. 
We identify two common concept drift mechanisms, the \ac{ddm} \cite{gama2004learning} and \ac{pht} \cite{page1954continuous}, as candidate methods for detecting the need for online training of deep receivers with and without pilots, respectively. These existing methods are designed for arbitrary hard decision classifiers, while receivers often employ soft probabilistic outputs (used for, e.g., log likelihood ratios). Thus, we propose a dedicated drift detection scheme for deep receivers based on the Hotelling test~\cite{hotelling1992generalization}, that is tailored to identify when a deep receiver needs to be re-trained by observing its soft outputs. 

Next, we focus on modular deep receivers, designed by augmenting conventional receiver processing algorithms with compact \acp{dnn} as a form of model-based deep learning~\cite{shlezinger2022model}. We leverage the interpretable modular architecture of such deep receivers, and particularly the fact that each sub-module is assigned with a concrete functionality, to support asynchronous learning of only the necessary sub-modules, rather than the entire receiver. To that aim, we extend our concept drift approach into \emph{modular drift detection}. There, the detector observes different sub-modules based on their designated interpretable meaning, such that  each functionality may be adapted asynchronously while the others are kept unchanged. 
We numerically demonstrate the gains of incorporating our  modular concept drift mechanisms, in  reducing the amount of re-training and its  burden compared with synchronous online training, without notably degrading performance.

    

Our main contributions are summarized as follows:
\begin{itemize}
    \item \textbf{Asynchronous online learning via drift detection:}
We identify the need for asynchronous online learning, stemming from the observation that not every variation between train and test channel necessities adaptation of deep receivers. We propose adopting the \ac{ml} framework of concept drift detection~\cite{bayram2022concept} as providing means for enabling asynchronous online learning of deep receivers.  
    \item \textbf{Soft-probabilities drift detection mechanism:} We adapt common drift detection algorithms to the domain of deep receivers, and introduce a novel soft-probabilities drift detection mechanism based on the Hotelling test \cite{hotelling1992generalization}. Our numerical results and analytical analysis demonstrate that complexity overhead can be greatly reduced, as compared to an synchronous online learning, by employing these detectors, while the loss in accuracy is minimal.
    \item \textbf{Modular drift detection:} We formulate drift detection mechanisms for interpretable sub-modules of model-based deep receivers \cite{shlezinger2020model,shlezinger2022model}. We propose to only re-train specific modules of the receiver architecture that need adaptation due to temporal variations of the channel, and provide a complexity analysis characterizing the savings of modular drift detection. Our numerical results validate the additional savings to computational complexity, again with negligible loss to accuracy.  
    \item \textbf{Extensive experimentation:} We extensively evaluate asynchronous online training using our proposed schemes for both \ac{siso} and \ac{mimo} systems, considering  different time-varying channel profiles: a single-user synthetic bursty channel, a multiple-users synthetic bursty channel, and COST 2100 \cite{liu2012cost} channels. The  findings of our experiments support the validity of our proposed approach.
\end{itemize}

The rest of this paper is organized as follows: 
In Section~\ref{sec:background} we review the system model. The drift detection methods are proposed for arbitrary deep receivers in Section~\ref{sec:soft_modular_drift_detection}, while Section~\ref{sec:modular} extends our approach to exploiting modular deep receiver architectures. Section~\ref{sec:Simulation} presents our numerical study, while Section~\ref{sec:conclusion} concludes the paper.

 Throughout the paper, we use boldface letters for vectors, e.g., ${\myVec{x}}$. 
 Calligraphic letters, such as $\mySet{X}$, are used for sets, with $|\mySet{X}|$ being the cardinality of $\mySet{X}$, and $\emptyset$ denoting the empty set.
\section{System Model}
\label{sec:background}

 In this section we detail the considered system model for asynchronous online learning. We first describe the block-fading communication system under study in Subsection~\ref{subsec:system_model}, and   review hard and soft-output receivers in Subsection~\ref{subsec:demodulators}. Then, we formulate the asynchronous online learning problem, which is the focus of this study,  in Subsection~\ref{subsec:problem}.

 \subsection{Time-Varying Channel Model}
 \label{subsec:system_model}
We consider a block-fading communication system in discrete-time. Each block is comprised of $\Blklen^{\rm tran}$ time instances, during which the channel parameters, denoted $\myVec{h}[t]$ for the $t$-th block, are constant. 
Let $\myVec{s}_i[t]\in\mySet{S}$ be a symbol transmitted from constellation $\mySet{S}$ at the $i$-th time instance of block $t$. The transmitted  block $\mys ^{\rm tran}[t]:= \{\myVec{s}_i[t]\}$ is divided into $\Blklen^{\rm pilot}$ known pilots,  denoted $\myVec{s}^{\rm pilot}[t]$, followed by   $\Blklen^{\rm info} = \Blklen^{\rm tran}-\Blklen^{\rm pilot}$  information symbols, denoted $\myVec{s}^{\rm info}[t]$. 

The channel output at time  $i$ is denoted  $\myVec{y}_i[t] \in \mySet{Y}$, and the received block is $\myVec{y}^{\rm rec}[t]:= \{\myVec{y}_i[t]\}$. Similarly to the channel input, the channel output can be separated by the receiver into its pilot and information parts denoted $\myVec{y}^{\rm pilot}[t]$ and $\myVec{y}^{\rm info}[t]$, respectively. 
We consider a generic channel model, determined by a conditional distribution  parameterized by  $\myVec{h}[t]$,
\begin{align}
    \label{eq:general_channel_mapping}
    \myVec{y}_i[t] \sim P_{\myVec{h}[t]}(\myVec{y}_i[t]|\myVec{s}_i[t]).
\end{align}
In~\eqref{eq:general_channel_mapping}, the channel model is subject to the (possibly unknown) conditional  distribution $P_{\myVec{h}[t]}(\cdot|\cdot)$,  that depends on the current channel parameters $\myVec{h}[t]$.

 \subsection{Deep Receivers}
 \label{subsec:demodulators}
The receiver  is aided by a  \ac{dnn}, which is comprised of $M$ distinct sub-modules. These modules can represent, e.g., parameterized iterations, as in deep unfolded optimizers~\cite{shlezinger2022model,balatsoukas2019deep}, or user-wise learned modules as in~\cite{shlezinger2019deepsic,van2022deep}. We use $\Weights^m[t]$ to denote parameters of the $m$th sub-module at block $t$, with $m\in \{1,\ldots,M\}:=\mySet{M}$, which are stacked into the overall \ac{dnn} parameter vector $\Weights[t] = [\Weights^1[t], \ldots, \Weights^M[t]]$. This formulation clearly accommodates non-modular \ac{dnn} architectures, where one can only associate their output with a concrete soft estimate, by setting $M=1$.

We focus on {\em soft-output receivers}, which map a channel output $\myVec{y}$ into an estimate of the conditional distribution of  $\myVec{s}$, denoted $\hat{P}_{\Weights[t]}\left(\myVec{s}|\myVec{y} \right)$. 
During information transmission, the soft estimate is used to provide a hard estimate denoted $\hat{\myVec{s}}_i[t]$, either by downstream processing (as in, e.g., \cite{shlezinger2019viterbinet,van2022deep}), or by taking the maximum a-posteriori probability rule~\cite{shlezinger2019deepsic}.

When applying {\em online learning} on block $t$, the receiver updates $\Weights[t]$ using the pilots data  $\mySet{Q}[t]:=\{\myVec{y}_i[t],\myVec{s}_i[t]\}_{i= 1}^{\Blklen ^{\rm pilot}}$. Training follows standard  \ac{sgd}-based learning with the  cross-entropy loss, seeking the solution to
 \begin{align}
\label{eq:loss}
  \mathop{\arg \max}_{\Weights[t]}  \sum_{(\myVec{y}_i, \myVec{s}_i) \in\mySet{Q}[t] }\log \hat{P}_{\Weights[t]}\left(\myVec{s}_{i}|\myVec{y}_{i} \right),
\end{align}
After training, the receiver recovers the information symbols from $\myVec{y}^{\rm info}[t]$ using its \ac{dnn}-aided  mapping. 
%

 \subsection{Problem Formulation}
 \label{subsec:problem}
Online training is typically studied assuming $(i)$ training of the entire set of parameters $\Weights[t]$ based on \eqref{eq:loss}; and $(ii)$ that such training is done on each coherence interval, i.e., at every block. We refer to this form of adaptation, which is clearly computationally extensive, as {\em synchronous online learning}. 
Here, we propose to alleviate the computational burden of online learning by adopting an asynchronous approach.

\begin{figure*}
    \centering
\includegraphics[width=\linewidth]{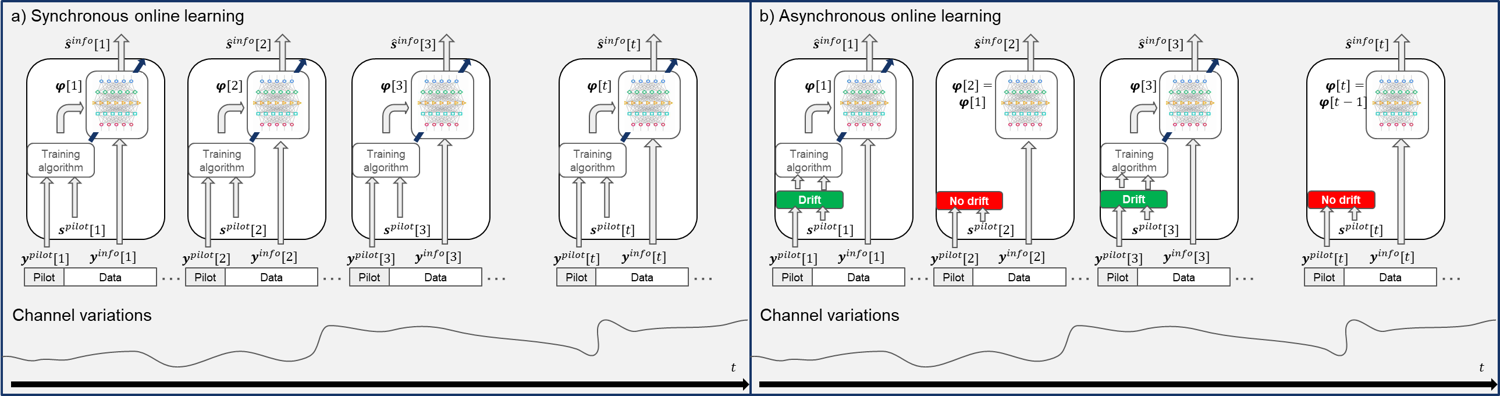}
    \caption{Schematic illustration of $(a)$ synchronous online learning compared to $(b)$ asynchronous online learning. }
    \label{fig:SyncVsAsync}
\end{figure*}

In {\em asynchronous online learning}, the receiver is allowed to decide on each block $t$: 
$(i)$ whether to apply training; and
$(ii)$ which sub-modules of $\Weights[t]$ to train.
To formulate this mathematically, we introduce the set $\TrainSet[t] \subseteq \mySet{M}$, whose elements represent the modules that are to be trained on block $t$. Setting $\TrainSet[t] = \emptyset$ indicates that no online learning is done. The operation of asynchronous online learning compared with its conventional synchronous counterpart is illustrated in Fig.~\ref{fig:SyncVsAsync}.

The setting of $\TrainSet[t]$ should aim at minimizing the \ac{ser} over the information symbols, while limiting the overall number of re-training operations. Over a horizon of $T$ blocks, this is formulated as
\begin{align}
\label{eqn:Problem} 
    	&\min \frac{1}{T \cdot \Blklen^{\rm info}}\sum_{t=1}^{T}\sum_{i= \Blklen^{\rm pilot} + 1}^{\Blklen^{\rm tran}} \Pr\left( \hat{\myVec{s}}_i[t] \neq  \myVec{s}_i^{\rm info}[t] \right)  \\
     &\text{ {subject to} } \sum_{t=1}^{T} |\TrainSet[t]| \leq C, \notag
\end{align}
where $C\ll T$  constrains the amount of re-training. If the threshold $C$ is too low, the receiver may suffer from performance degradation, while if it is too large the complexity gain of the scheme is minimal over  synchronous online training. Thus, this hyperparameter should be carefully chosen based on the varying nature of the test scenario and the desired gains in complexity. This is further explored in Subsection~\ref{subsec:complexity_analysis}. Also, while analytically evaluating the probability $\Pr\left( \hat{\myVec{s}}_i[t] \neq  \myVec{s}_i^{\rm info}[t] \right)$ is challenging for the considered complex channels models and when using \ac{dnn}-aided receivers, this quantity can be evaluated empirically (see Section~\ref{sec:Simulation}).

\begin{figure}[!t]
\removelatexerror
  \begin{algorithm}[H]
    \caption{Supervised \ac{ddm} Algorithm }
    \label{alg:ddm}
    \SetAlgoLined
    \SetKwInOut{Input}{Input}
    \Input{Pilots $\mySet{Q}[t]$; \ac{dnn} $\Weights[t]$; Threshold $\lambda$;
    \newline Previous  moments $\mu[t-1],{\sigma}[t-1]$; 
    \newline Forgetting parameter $\beta$;
    }
    \SetKwProg{TheDDMAlgorithm}{DDM}{}{}
    \TheDDMAlgorithm{($\mySet{M}$)}{
    Use \ac{dnn} modules $\mySet{M}$ to compute $\hat{\myVec{s}}^{\rm pilot}[t]$\\
    \label{line:u0}

    Calculate $\mu[t]$ and $\sigma[t]$ via \eqref{eqn:DDMmoments} \\

    \uIf{$\mu[t] + \sigma[t] > \mu[t-1] + \lambda \cdot \sigma[t-1]$ }
    {
     \KwRet{$\TrainSet[t]=\mySet{M}$}
    }
    \uElse{
    $\mu[t] \leftarrow \beta \mu[t] + (1-\beta)\mu[t-1]$ \\
    $\sigma[t] \leftarrow \beta \sigma[t] + (1-\beta)\sigma[t-1]$ \\
    \KwRet{$\TrainSet[t]=\emptyset$}
    }


    


    
    
  }
  \end{algorithm}
\end{figure}

\section{Unstructured Asynchronous Online Learning}
\label{sec:soft_modular_drift_detection}
We commence by tackling \eqref{eqn:Problem} without accounting for possible structures in the deep receiver architecture. 
Accordingly, in this section we restrict the optimized set $\TrainSet[t]$ to take binary values, i.e., it can either be $\emptyset$ (indicating no adaptation) or $\mySet{M}$ (thus training the entire architecture). This form of unstructured asynchronous learning is used as a basis for modular asynchronous online learning, detailed in Section~\ref{sec:modular}.

As we do not account here for the \ac{dnn} structure, we focus on introducing drift detection mechanisms that enable asynchronous online learning for deep receivers. We first adapt existing detectors from the \ac{ml} literature based on the  supervised \ac{ddm} \cite{gama2004learning} and unsupervised \ac{pht} \cite{page1954continuous} in Subsection~\ref{subsec:ddm}. We then propose a soft-output mechanism in Subsection~\ref{subsec:method}, and discuss the considered methods in Subsection \ref{subsec:discussion}. 



\subsection{Conventional Detectors}
\label{subsec:ddm}
Various methods are proposed in the \ac{ml} literature for drift detection, mostly focusing on \ac{ml} models trained for hard classification. Typically, such schemes are divided into {\em supervised} and {\em unsupervised} settings. Using the notations of Subsection~\ref{subsec:problem}, supervised detectors determine $\TrainSet[t]$ based on $\Weights[t-1]$ and the observed pilots  $\mySet{Q}[t]$, while {\em unsupervised} mechanisms use only the channel output, i.e., $\myVec{y}[t]$, in addition to $\Weights[t-1]$. We next adapt two candidate methods for our setting of unstructured asynchronous online learning of deep receivers: the supervised \ac{ddm} and the unsupervised \ac{pht}. 

\subsubsection{\ac{ddm}} 
One of the most popular supervised drift detector is the \ac{ddm}. It is based on the premise that when the distribution of stream-based data changes, the classification error  increases. 
\ac{ddm} is obtained by assuming that the average number of errors obeys a (scaled) binomial distribution \cite{gama2004learning}. Retraining is declared when the empirical standard deviation above the empirical mean exceeds a confidence interval determined from previous blocks.

Adapting the \ac{ddm} to our setting, the empirical mean  $\mu[t]$ is obtained as the error rate over the pilot blocks. Letting $1_{\mySet{A}}$ be the indicator  of event $\mySet{A}$, we compute
\begin{subequations}
    \label{eqn:DDMmoments}
\begin{equation}
\label{eqn:bernoulli_mu}
    \mu[t] = \frac{1}{B^{\rm pilot}}\sum_{i=1}^{B^{\rm pilot}}1_{\hat{\myVec{s}}_i[t] \neq \myVec{s}_i[t]}.
\end{equation}
The empirical standard deviation $\sigma[t]$ is computed by the binomial standard deviation as
\begin{equation}
\label{eqn:bernoulli_std}
	\sigma[t] = \sqrt{\frac{\mu[t](1-\mu[t])}{B^{\rm pilot}}}.
\end{equation} 
\end{subequations}

Using the empirical estimates in \eqref{eqn:DDMmoments},  \ac{ddm} checks whether the statistic $\mu[t] + \sigma[t]$ exceeds the confidence interval $[0,\mu[t-1] + \lambda\sigma[t-1]]$, indicating a significant error rate that requires retraining, where $\lambda$ is a  constant representing a configurable threshold. If no re-training is needed, the empirical moments in \eqref{eqn:DDMmoments} are combined along with  $\mu[t-1],\sigma[t-1]$ with forgetting factor $\beta\in[0,1]$ to be used for the next block.
The procedure is summarized as Algorithm~\ref{alg:ddm}.



\subsubsection{\ac{pht}} 
A common unsupervised drift detection algorithm is based on the \ac{pht}.
This test detects whether changes occur in the channel output distribution, by measuring  deviations in the first-order moments.
Unlike \ac{ddm}, which declares re-training based on the output of  the \ac{dnn},  \ac{pht} \cite{page1954continuous} is invariant of the \ac{dnn}, and is effectively an online change point detection algorithm often used for drift detection~\cite{aminikhanghahi2017survey}.   
 Accordingly, it is not restricted to pilots, and can  use the entire transmitted block. 

To detect changes, the cumulative mean $\myVec{\mu}[t]$ is calculated  via
\begin{equation}
\label{eqn:cumsum}
	\myVec{\mu}[t] = \frac{\beta}{\Blklen^{\rm tran
 }}\sum_{i=1}^{\Blklen^{\rm tran}} \myVec{y}^{\rm mag}_i[t] + (1-\beta)\myVec{\mu}[t-1],
\end{equation}
with forgetting factor $\beta\in[0,1]$ and $\myVec{y}^{\rm mag}[t] \equiv \lVert\myVec{y}^{\rm rec}[t]\rVert$.
Then, the aggregated distance between the channel outputs and $\myVec{\mu}[t]$ is calculated as 
\begin{equation}
\label{eqn:distance}
	d[t] = \max\Big\{0,\sum_{i=1}^{\Blklen^{\rm tran}} (\lVert \myVec{y}^{\rm mag}_i[t] - \myVec{\mu}[t] \rVert - \delta)\Big\},
\end{equation} 
with $\delta$ being the magnitude change factor. 
The drift detection  statistics is the difference between $d[t]$ and $d[t-1]$, i.e., re-training is declared when it exceeds a threshold $\lambda$. This process is described as Algorithm~\ref{alg:pht}.

\begin{figure}[!t]
\removelatexerror
  \begin{algorithm}[H]
    \caption{Unsupervised \ac{pht} Algorithm }
    \label{alg:pht}
    \SetAlgoLined
    \SetKwInOut{Input}{Input}
    \Input{Received  $\myVec{y}^{\rm rec}[t]$;  Threshold $\lambda$ 
    \newline 
    Previous statistics $\myVec{\mu}[t-1],d[t-1]$; 
    \newline 
    Forgetting parameter $\beta$; Change factor $\delta$; 
    }  
    \SetKwProg{ThePHTAlgorithm}{PHT}{}{}
    \ThePHTAlgorithm{$(\mySet{M})$}{
    

    \label{line:u0}
    Calculate $\myVec{\mu}[t]$ from \eqref{eqn:cumsum}\\

    Calculate $d[t]$ from \eqref{eqn:distance}\\

    \uIf{$|d[t] - d[t-1]| > \lambda$}
    {
  {  $\myVec{\mu}[t] \leftarrow \frac{1}{\Blklen^{\rm tran
 }}\sum_{i=1}^{\Blklen^{\rm tran}} \lVert\myVec{y}^{\rm rec}_i[t]\rVert$ }\\
     \KwRet{$\TrainSet[t]=\mySet{M}$}
    }
    \uElse{
    \KwRet{$\TrainSet[t]=\emptyset$}\label{line:return1} \
  }
  }
  \end{algorithm}
\end{figure}

\subsection{Proposed Soft-Output Drift Detectors}
\label{subsec:method}
 \ac{ddm} and \ac{pht} are well-established drift detectors in the literature on \ac{ml}  for online streaming data. 
However, in our context of deep receivers,  \ac{ddm}  views the receiver as a black-box hard decision classifier, thus not exploiting its soft-outputs; \ac{pht} only observes the channel output, and is designed to identify when its distribution changes rather than when the receiver  needs re-training. Accordingly, we  derive drift detection methods that leverage the presence of pilots, while exploiting the soft-outputs of deep receivers.  

The rationale for our method is based on the fact that changing $\myVec{h}[t]$ affects the  distribution  \eqref{eq:general_channel_mapping}, namely, $P_{\myVec{h}[t]}(\myVec{y}_i[t]|\myVec{s}_i[t]=s)$ conditioned on the transmitted  $s\in \mySet{S}$. Accordingly, following~\cite{shlezinger2021learned}, such changes are expected to reflect on the receiver's output corresponding to that symbol,  i.e., $\hat{P}_{\Weights[t]}\left(\myVec{s}_{i}[t]=s|\myVec{y}_{i}[t] \right)$. We thus treat the values of $\hat{P}_{\Weights[t]}\left(\myVec{s}_{i}[t]=s|\myVec{y}_{i}[t] \right)$ when the transmitted pilot is $s\in\mySet{S}$ as {\em independent realizations} of a random variable. Note that in a single block, where the channel distribution remains approximately constant, these soft outputs are identically distributed.

Based on this representation, we formulate two soft-output oriented drift detection schemes: 
The first, termed {\em Posterior-Based Drift Detection}, utilizes only the mean of the distribution and is thus very simple to implement; 
the second extends the posterior-based detector to improve detection accuracy at the cost of some complexity increase. This is achieved by  employing the empirical covariance as well using  Hotelling's two-sample $t$-squared test  \cite{hotelling1992generalization}, and thus the test is coined {\em Hotelling Drift Detection}.


\subsubsection{Posterior-Based Drift Detector}
For each $s\in\mySet{S}$,  let $\mySet{I}^s[t]$ be the time indices where $s$ is used as pilot, i.e., 
\begin{equation*}
\mySet{I}^s[t] := \{i\in 1,\ldots,B^{\rm pilot}| \myVec{s}_i[t]=s\}.     
\end{equation*}
Next, we gather the soft outputs corresponding to $s$, i.e., $\{\hat{P}_{\Weights[t]}\left(s|\myVec{y}_{i}[t] \right)\}_{i\in \mySet{I}^s[t]}$, which we treat as independent realizations of a random variable. 

In the first method, we only consider the empirical mean $\mu^s$ of the posterior for each symbol $s$, i.e., 
\begin{equation}
\label{eqn:mu_s}
    \mu^s[t] = \frac{1}{|\mySet{I}^s[t] |}\sum_{i\in \mySet{I}^s[t]}\hat{P}_{\Weights[t]}\left(s|\myVec{y}_{i}[t] \right).
\end{equation}
Then, we compute the average of the empirical means across all the symbols as
\begin{equation}
\label{eqn:posterior_mu}
    \mu[t] = \sum_{s\in \mySet{S}} \frac{|\mySet{I}^s[t] |}{\Blklen^{\rm pilot}} \mu^s[t].
\end{equation}
If this value falls below a probability threshold $\lambda$, then the receiver is deemed unsure of the probabilities and re-training occurs.

The resulting algorithm, summarized in full as Algorithm \ref{alg:posterior_proposed}, is simple to implement, as it requires only an estimate of the first-order moment of the posterior. The test can be extended to also account for second-order moments, as detailed next.

\begin{figure}[!t]
\removelatexerror
  \begin{algorithm}[H]
    \caption{Posterior-Based Drift Detection}
    \label{alg:posterior_proposed}
    \SetAlgoLined
    \SetKwInOut{Input}{Input}        
    \Input{Pilots $\mySet{Q}[t]$; \ac{dnn} $\Weights[t]$; Threshold $\lambda$;
    \newline Previous  means $\{\mu^s[t-1]\}$; 
    \newline Forgetting parameter $\beta$;
    }  
    \SetKwProg{TheProposedAlgorithm}{Posterior Drift Detection}{}{}
    \TheProposedAlgorithm{($\mySet{M}$)}{
    Compute $\{\hat{P}_{\{\Weights_m[t]\}_{m\in\mySet{M}}}\left(\myVec{s}|\myVec{y} \right)\}_{(\myVec{s},\myVec{y})\in \mySet{Q}[t]}$ \\
    
    \For{$s\in\mathcal{S}$}{ 
    
            Calculate $\mu^s[t]$ via \eqref{eqn:mu_s}\\
        }

    Calculate $\mu[t]$ from \eqref{eqn:posterior_mu}\\


    \uIf{$\mu[t] < \lambda$ }
    {
     \KwRet{$\TrainSet[t]=\mySet{M}$}
    }
    \uElse{
    $\mu^s[t] \leftarrow \beta \mu^s[t] + (1-\beta)\mu^s[t-1]$, $\forall s\in\mySet{S}$ \\
    \KwRet{$\TrainSet[t]=\emptyset$}
    }
  }
  \end{algorithm}
\end{figure}

 \subsubsection{Hotelling Drift Detector}
 While accounting merely for the estimated mean of the posterior is simple to implement, it does not consider the variations in its value observed within the observed pilots. To account for these statistics as well, we propose a drift detector that is based on Hotelling's test. The proposed detector follows the same computation of the empirical mean as in \eqref{eqn:mu_s}, but also involves its empirical variance via 
\begin{equation}
\label{eqn:covariance_s}
 \sigma^s[t] =  \frac{1}{|\mySet{I}^s[t] | - 1} \sum_{i\in \mySet{I}^s[t]}\left( \hat{P}_{\Weights[t]}\left(s|\myVec{y}_{i}[t]\right) - \mu^s[t]  \right)^2. 
\end{equation}

 Hotelling's two-sample $t$-squared test  is an affine-invariant and asymptotically constant-false-alarm-rate detector, which is designed to detect a difference between the first-order moments of two compared distributions~\cite{hotelling1992generalization}. To that sake, it employs a normalized version of the squared difference between the respective empirical means, computed as  
\begin{equation}
	\hat{{\sigma}}^s[t] \!=\! \frac{(|\mySet{I}^s[t] |\!-\!1){\sigma}^s[t] \!+\! (|\mySet{I}^s[t-1] |\!-\!1){\sigma}^s[t-1]}{|\mySet{I}^s[t] | + |\mySet{I}^s[t-1] | - 2}. 
\end{equation} 
The Hotelling test statistic is given by 
\begin{equation}
\label{eqn:Hotelling}
    \mathcal{T}^s[t] = \frac{|\mySet{I}^s[t] | \cdot |\mySet{I}^s[t-1] |}{|\mySet{I}^s[t] | + |\mySet{I}^s[t-1] |} \cdot\frac{(\mu^s[t]-\mu^s[t-1])^2}{\hat{{\sigma}}^s[t]}. 
\end{equation}

We thus obtained an ensemble of $|\mySet{S}|$ test statistics, each taken from a possibly different number of realizations. Accordingly, the statistic  for detecting if re-training is needed is obtained by weighted averaging \eqref{eqn:Hotelling}  over $s\in\mySet{S}$, i.e., 
\begin{equation}
\label{eqn:t_final}
	\mathcal{T}[t] = \sum_{s\in\mathcal{S}} \frac{|\mySet{I}^s[t] |}{\Blklen^{\rm pilot}}  \mathcal{T}^s[t]. 
\end{equation} 
If $\mathcal{T}[t]$ deviates from a chosen threshold $\lambda$, then re-training is declared. 
The  procedure is summarized as Algorithm~\ref{alg:proposed}. 

For a single Hotelling test, the threshold can be determined from the asymptotic distribution of the statistic~\cite{hotelling1992generalization} (i.e., when the number of pilots with a given $s\in\mySet{S}$ is asymptotically large). This setting allows meeting a desired rate of 'false alarms', which in our case represents unnecessary re-training operations. However, in our ensemble test,  \eqref{eqn:t_final} is obtained by a weighted sum of correlated test statistics~\cite{peligrad1987central}, and we are particularly interested in settings where the number of pilots is limited and small. Accordingly, we set $\lambda$ by empirical trials, and leave the asymptotic analysis of \eqref{eqn:t_final} for future work.

\begin{figure}[!t]
\removelatexerror
  \begin{algorithm}[H]
    \caption{Soft-Output Hotelling Drift Detection}
    \label{alg:proposed}
    \SetAlgoLined
    \SetKwInOut{Input}{Input}        
    \Input{Pilots $\mySet{Q}[t]$; \ac{dnn} $\Weights[t]$; Threshold $\lambda$;
    \newline Previous  moments $\{\mu^s[t-1],{\sigma}^s[t-1]\}$; 
    \newline Forgetting parameter $\beta$;
    }  
    \SetKwProg{TheProposedAlgorithm}{Hotelling Drift Detection}{}{}
    \TheProposedAlgorithm{($\mySet{M}$)}{
    Compute $\{\hat{P}_{\{\Weights_m[t]\}_{m\in\mySet{M}}}\left(\myVec{s}|\myVec{y} \right)\}_{(\myVec{s},\myVec{y})\in \mySet{Q}[t]}$ \\
    
    \For{$s\in\mathcal{S}$}{ 
    
            Calculate $\mu^s[t]$ and ${\sigma}^s[t]$ via \eqref{eqn:mu_s} and \eqref{eqn:covariance_s}\\

            Calculate $\mySet{T}^s[t]$ from \eqref{eqn:Hotelling}\\
        }

    Calculate $\mySet{T}[t]$ from \eqref{eqn:t_final}\\


    \uIf{$\mySet{T}[t] > \lambda$ }
    {
     \KwRet{$\TrainSet[t]=\mySet{M}$}
    }
    \uElse{
    $\mu^s[t] \leftarrow \beta \mu^s[t] + (1-\beta)\mu^s[t-1]$, $\forall s\in\mySet{S}$ \\
    $\sigma^s[t] \leftarrow \beta \sigma^s[t] + (1-\beta)\sigma^s[t-1]$, $\forall s\in\mySet{S}$ \\
    \KwRet{$\TrainSet[t]=\emptyset$}
    }
  }
  \end{algorithm}
\end{figure}

\subsection{Discussion}
\label{subsec:discussion}

The  detectors proposed in the previous subsections  allow asynchronous online learning while operating with reduced complexity. To provide a comparative discussion, we note that all methods track test statistics whose computation involves simple averaging procedures. 
In particular, \ac{ddm} (Algorithm~\ref{alg:ddm}) tracks merely a single scalar statistic, based on the error rate.  Even though its simplicity and minimal memory footprint are a great fit for low resource scenarios, this supervised method can potentially falter when the scenario exhibits gradual slow changes or the labeled data is limited. 
\ac{pht} tracks a test statistic for each input, and all are used to compute an additional distance metric, i.e., if the input is a $|\mySet{Y}|\times 1$ vector, then $|\mySet{Y}| + 1$ variables are tracked. Its operation does not require pilots, being  an unsupervised \ac{ml} approach. Yet, it only detects when the channel changes, which may not always be indicative of when re-training is needed due to performance deterioration. Moreover, these two methods, while being conventional drift detection techniques in the  \ac{ml} literature, are based on hard decision values, unable to exploit the outputs of soft receivers. 

Our proposed soft-input detectors formulated in Subsection~\ref{subsec:method}
exploit the soft probabilities of deep receivers. The  posterior-based detector achieves this by monitoring fluctuations in the mean of the predicted posterior for each distinct symbol, thus tracking $|\mySet{S}|$ variables. The Hotelling   detector also tracks second-order moments, i.e., $2|\mySet{S}|$ variables. Its additional complexity compared with the posterior detector allows achieving improved asynchronous learning, as numerically demonstrated in Section~\ref{sec:Simulation}.  Still, both approaches enable asynchronous online learning, and are  characterized by ease of implementation while requiring a negligible memory footprint.

The proposed soft detectors, as well as \ac{ddm}, are designed to identify fluctuations in the receiver and not in the channel, and are thus expected to lead to asynchronous online learning on the specific blocks where such adaptation is indeed required. This behavior is systematically observed in our numerical study in Section~\ref{sec:Simulation}. While the considered algorithms are {\em unstructured}, i.e., invariant of the modular architecture, and aim at detecting drifts in the overall \ac{dnn} comprised of $M$ modules (regardless of $M$), they can be extended to enable {\em modular} asynchronous learning, as detailed next. 



\section{Modular Asynchronous Online Learning}
\label{sec:modular}

Asynchronous online learning using the detectors detailed so far  aims at reducing the number of re-training times while maintaining the error rate as low as possible. Nonetheless, re-training of highly-parameterized deep receivers may still induce notable computational burden even after reducing the re-training times. In the following we focus on deep receivers that employ hybrid model-based/data-driven architectures. These architectures follow the processing chain of classic model-based receiver processing,  comprised of an interconnection of task-oriented modules, while promoting flexibility via the parameterization of each module. Such architectures tend to be less parameterized as compared to black-box end-to-end \acp{dnn}~\cite{farsad2020data}, and their modular architecture can be leveraged to facilitate learning more rapidly~\cite{raviv2022online} and with less data~\cite{raviv2023modular}. Here, we show that their modularity can also facilitate asynchronous online learning. 

To that end, we  extend the asynchronous online learning framework of Section~\ref{sec:soft_modular_drift_detection} to take into account the modular structure of such deep receivers, allowing for different levels of adaptation per each module. 
Our approach is formulated  in 
Subsection~\ref{subsec:modular_formulation}. 
We provide an analysis of the  complexity of modular asynchronous online learning in Subsection~\ref{subsec:complexity_analysis}, and conclude with a discussion in Subsection~\ref{subsec:modular_discussion}.

\subsection{Modular Online Learning}
\label{subsec:modular_formulation}
The formulation of the deep receiver in Subsection~\ref{subsec:system_model} considers  \acp{dnn} that can be viewed as an interconnection of $M$ modules. This formulation is quite general, as, e.g., one can treat each layer or residual block of a \ac{dnn} as a module. Here, we specifically focus on architectures where each layer has an {\em operational meaning}, in the sense that one can evaluate its performance based on its output features, and not just based on the  output of the overall \ac{dnn}. Such modular architectures naturally arise when designing \acp{dnn} via deep unfolding~\cite{shlezinger2022model}, which is extensively studied in the context of wireless communications~\cite{balatsoukas2019deep}, with candidate receiver architectures including the DeepSIC symbol detector~\cite{shlezinger2019deepsic} and the WBP decoder~\cite{nachmani2018deep}.

In modular deep receivers, each module is designed to fulfill specific functions within the communication receiver.  Notably, certain functionalities necessitate rapid adaptation, while others remain unchanged over a longer duration. 
We focus on settings where each module can be assessed (and thus trained) separately. For instance, unfolded modular architectures as in~\cite{shlezinger2019deepsic,nachmani2018deep,van2022deep}  have each module produce a soft estimate of a subset of the transmitted data, i.e., the $m$th module output at block $t$ is a soft estimate $\hat{P}_{\Weights_m[t]}\left(\myVec{s}|\myVec{y} \right)$. Accordingly, each module parameters $\Weights_m[t]$ can be online trained to approach 
 \begin{equation}
\label{eq:dynamic_only_loss}
  \mathop{\arg \max}_{\Weights^m[t]}  \sum_{(\myVec{y}_i, \myVec{s}_i) \in\mySet{Q}[t] }\log \hat{P}_{\Weights^m[t]}\left(\myVec{s}_{i}|\myVec{y}_{i} \right).
\end{equation}

 Modular asynchronous online learning leverages this interpretable structure to further facilitate re-training, utilizing drift detection mechanisms to identify {\em which modules} are to be retrained. In this case, the set of modules requiring adaptation at block $t$, denoted $\TrainSet[t]$, can be any subset of $\mySet{M}$. The modular asynchronous learning thus applies drift detection via any of the mechanisms detailed in Section~\ref{sec:soft_modular_drift_detection} for each module separately, and adapts only those identified as requiring re-training. 
 The resulting procedure is summarized as Algorithm~\ref{alg:Asynchronous} (formulated for supervised pilot-aided drift detectors), where the method {\bf Drift Detect} in Step~\ref{stp:detect} can be replaced with any of Algorithms~\ref{alg:ddm}-\ref{alg:proposed}.
 The constraint on the overall number of re-training modules in \eqref{eqn:Problem} is enforced by proper setting of the threshold $\lambda$. 
 
 It is noted that the online adaptation in Step~\ref{stp:train} does not specify how the identified modules in $\TrainSet[t]$ are re-trained. In general, they can be jointly adapted based on the overall system output as in \eqref{eq:loss}. Alternatively, as also done in our experimental study in Section~\ref{sec:Simulation}, the modules can be trained sequentially (see \cite{shlezinger2019deepsic}), with each module adapted based on \eqref{eq:dynamic_only_loss}. 
 Moreover, while Step~\ref{stp:detect} is formulated as examining every module separately, it can be readily extended to examining distinct subset of modules by replacing $\{m\}$ with the corresponding subset of $\mySet{M}$.

\begin{figure}[!t]
\removelatexerror
  \begin{algorithm}[H]
    \caption{Modular Asynchronous Online Learning}
    \label{alg:Asynchronous}
    \SetAlgoLined
     \SetKwInOut{Input}{Input}        
     \Input{Pilots $\mySet{Q}[t]$; \ac{dnn} $\Weights[t]$; 
     }  
    \SetKwProg{TheProposedAlgorithm}{Asynchronous Online Learning}{}{}
    \TheProposedAlgorithm{}
   {
    Set $\TrainSet[t] = \emptyset$\\
    \For{$m\in\mathcal{M}$}{     
            $\TrainSet[t] \leftarrow\TrainSet[t]  \cup$ {\bf Drift Detection}$(\{m\})$ \label{stp:detect}\\
        }

    \uIf{$\TrainSet[t] \neq \emptyset$}{     
        Online adapt $\{\Weights_m[t]\}_{m\in\TrainSet[t]}$ \label{stp:train}\\
    } 
  }
  \end{algorithm}
\end{figure}

\subsection{Complexity Analysis}
\label{subsec:complexity_analysis}

Asynchronous modular online learning can greatly facilitate the operation of deep receivers in dynamic wireless channels. Given the considerable resource and latency expenditure associated with re-training, it is imperative to limit the number of re-training operations.  In the following, we  analyze the computational complexity of asynchronous online training (both modular and unstructured) with drift detection and compare it to synchronous online training. 

In our analysis we introduce the following symbols:
\begin{itemize}
    \item {\em Detection complexity} $\kappa_{\rm D}$, representing the effort in identifying a drift via the methods detailed in Section~\ref{sec:soft_modular_drift_detection}. As discussed in Subsection~\ref{subsec:discussion}, once the \ac{dnn} is applied, Algorithms~\ref{alg:ddm}-\ref{alg:proposed} do not depend on the \ac{dnn} or the number of its modules. Specifically,   $\kappa_{\rm d}$ is constant, depending only on the input (for PHT) or output (for soft-output detectors) dimensions. 
    \item {\em Training complexity} $\kappa_{\rm T}(\Weights)$, representing the computational effort in training a \ac{dnn} with parameters $\Weights$. For standard \ac{sgd}-based training, complexity scales linearly with the number of trainable parameters~\cite[Ch. 11]{zadeh2015cme}, and thus we write $\kappa_{\rm T}(\Weights)= \sum_{m=1}^M\kappa_{\rm T}(\Weights_m)$.  
\end{itemize}
For simplicity (and also corresponding to the common practice in unfolded architectures \cite{shlezinger2019deepsic, nachmani2018deep}), we assume that each module has the same number of parameters, such that $\kappa_{\rm T}(\Weights_m) = \frac{1}{M}\kappa_{\rm T}(\Weights)$ for each $m\in\mySet{M}$.

Using the above notations, we can characterize the computational effort due to online learning, as stated in the following proposition:
\begin{proposition}
	\label{pro:Complexity}
	The excessive computational complexity of modular asynchronous online learning via Algorithm~\ref{alg:Asynchronous} on  each block of index $t$ is given by
	\begin{equation}
	\label{eqn:Complexity}
	C_{\rm Modl}(M)  = M \cdot \kappa_{\rm d} +\frac{\kappa_{\rm T}(\Weights)}{M}\sum_{m=1}^{M}\Pr\left(m \in \TrainSet[t]\right).
	\end{equation}
\end{proposition}

\begin{IEEEproof}
	The first summand in \eqref{eqn:Complexity} stems from the fact that each of the $M$ modules is inspected (Step~\ref{stp:detect} of Algorithm~\ref{alg:Asynchronous}), which is thus given by $M\cdot \kappa_{\rm d} $. 
	The second represents the complexity of online training (Step~\ref{stp:train} of Algorithm~\ref{alg:Asynchronous}), in which each module of index $m$ is re-trained at complexity $\kappa_{\rm T}(\Weights_m) = \frac{1}{M}\kappa_{\rm T}(\Weights)$ whenever $m \in \TrainSet[t]$. 
\end{IEEEproof}

 Proposition~\ref{pro:Complexity} is stated for modular asynchronous online learning. Nonetheless, we note that this form of asynchronous learning specializes  unstructured asynchronous learning by setting $M=1$, and by observing that the detection complexity $\kappa_{\rm d}$ is invariant of the \ac{dnn} size. Consequently, we can use \eqref{eqn:Complexity} to characterize the corresponding complexity, as stated in the following corollaries:
 \begin{corollary}
 	\label{cor:ComplexityUnst}
 		The excessive computational complexity of unstructured asynchronous online learning on  each block of index $t$ is given by
 	\begin{equation}
 	\label{eqn:ComplexityUnst}
 	C_{\rm Unst}  = C_{\rm Modl}(1)= \kappa_{\rm d} +\Pr\left( \TrainSet[t] \neq \emptyset\right)\kappa_{\rm T}(\Weights).
 	\end{equation}
 \end{corollary}

The characterizations of the excessive complexity of the considered forms of asynchronous online allow one to compare the complexity savings of leveraging modularity, as well as that of asynchronous online learning compared to its synchronous counterpart, whose excessive complexity is  
 	\begin{equation}
 	\label{eqn:ComplexitySync}
 	C_{\rm Sync}  =  \kappa_{\rm T}(\Weights).
 	\end{equation} 
 In particular, comparing \eqref{eqn:Complexity} with \eqref{eqn:ComplexitySync}, reveals that the reduction in computation of asynchronous over synchronous online learning is given by
 \begin{align}
 \label{eqn:CompSave}
 \frac{C_{\rm Modl}(M)}{C_{\rm Sync}} 
 &= M  \frac{\kappa_{\rm d}}{\kappa_{\rm T}(\Weights)} +\frac{1}{M}\sum_{m=1}^{M}\Pr\left(m \in \TrainSet[t]\right).
 \end{align} 
 
 We note that, as discussed in Subsection~\ref{subsec:discussion} the computational effort of detection tests via Algorithms~\ref{alg:ddm}-\ref{alg:proposed} is negligible compared to that of training a \ac{dnn}. Consequently, 
 \begin{equation*}
 M\frac{\kappa_{\rm d}}{\kappa_{\rm T}(\Weights)} =\frac{\kappa_{\rm d}}{\kappa_{\rm T}(\Weights_m)}  \approx 0, 
 \end{equation*}
 for any \ac{dnn} module $m\in\mySet{M}$. 
  Accordingly, the complexity reduction of asynchronous operation can be approximated as
  \begin{align}
 \label{eqn:CompSave2}
 \frac{C_{\rm Modl}(M)}{C_{\rm Sync}} 
 &\approx  \frac{1}{M}\sum_{m=1}^{M}\Pr\left(m \in \TrainSet[t]\right) \leq 1.
 \end{align}
 By \eqref{eqn:CompSave2}, asynchronous online learning is expected to reduce the computational effort compared to synchronous online learning, where the reduction becomes more substantial when the probability of detecting drifts reduces.

\subsection{Discussion}
\label{subsec:modular_discussion}
 Extending asynchronous online learning to operate in a modular fashion
 exploits the interpretable structure of hybrid model-based/data-driven deep receivers. This is achieved by applying  drift detection to the subset of relevant modules in the overall  network, i.e., using the mechanisms as in Algorithms~\ref{alg:ddm}-\ref{alg:proposed} to adapt  less parameters. 
 Our complexity analysis indicates that asynchronous operation is expected to alleviate some of the computational burden of synchronous online learning. The saving ratio, given in \eqref{eqn:CompSave2}, depends on both the detection threshold $\lambda$, as well as on the nature of the variations of the underlying channel. In Section~\ref{sec:Simulation}, we empirically demonstrate that substantial reductions in computational effort, i.e., in the amount of parameters re-trained and re-training times, are achieved with only a minor performance loss compared to synchronous online learning in various channel variation profiles. 
 
 Comparing Corollary~\ref{cor:ComplexityUnst} and Proposition~\ref{pro:Complexity} indicates that leveraging modular architectures is expected to be most beneficial in settings where $(i)$ the detection compleixty $\kappa_{\rm d}$ is negligble compared to the training complexity (which is the expected case, as discussed above); and $(ii)$ the channel variations typically require adapting only subsets of the overall architecture. The latter is expected to occur when the variations affect only certain modules, e.g., when only part of the users are mobile in an uplink \ac{mimo} setup. In the extreme settings where only a single module is to be adapted, such that the events $\{m\in\TrainSet[t]\}$ are mutually exclusive, then 
 \begin{align*}
     \Pr\left(m \in \TrainSet[t]\right) &= \Pr \left(\cup_{m=1}^M m \in \TrainSet[t] \right) \notag \\ 
     &= \sum_{m=1}^M\Pr\left(m \in \TrainSet[t]\right),
 \end{align*} 
 and a complexity reduction by a factor of $M$ is achieved due to the modular operation.

It is noted that our design focuses on settings where the \ac{dnn} can be divided into modules that can be evaluated separately, and for which one can expect certain channel variations to require adapting only subsets of the architecture. One can potentially consider using this approach in black-box \acp{dnn} by viewing sets of layers or residual blocks as modules, while evaluating drifts based on the overall outputs.  
In a broader sense, our proposed scheme could be combined with other solutions that facilitate adaptation in dynamic communication scenarios with scarce labeled data~\cite{raviv2023adaptive}. For example, data augmentations \cite{almahairi2018augmented,raviv2022data} that exploit the unique invariance properties of communication systems could be used to enrich the small pilot dataset. Alternatively, meta-learning training \cite{park2020learning,raviv2022online} may allow faster convergence due to a module-wise inductive bias that is valid across many different channels. Similarly, the recent identification of Bayesian deep learning as facilitating training deep receivers from few pilots~\cite{cohen2022bayesian,raviv2023modular} indicate the need for new forms of concept drift detectors. We leave these extensions to future works.

\section{Numerical Evaluations}
\label{sec:Simulation}

In this section we numerically evaluate the asynchronous online learning framework across various scenarios for different \ac{dnn}-aided receiver architectures\footnote{The source code  is available at \url{https://github.com/nicoleva/modular-concept-drift-for-receivers}}. The details of our experimental setups are provided in Subsection~\ref{ssec:SimSetup}. We present our numerical evaluations for \ac{siso} and \ac{mimo} channels in Subsections~\ref{ssec:SimSISO} and \ref{ssec:SimMIMO}, respectively.

\subsection{Experimental Setup}
\label{ssec:SimSetup}
We evaluate asynchronous online learning over two main channels:
\subsubsection{Finite-Memory SISO Channels}
\begin{figure}
  \centering
  \subfigure[SISO COST2100]
    {\includegraphics[width=0.8\linewidth]{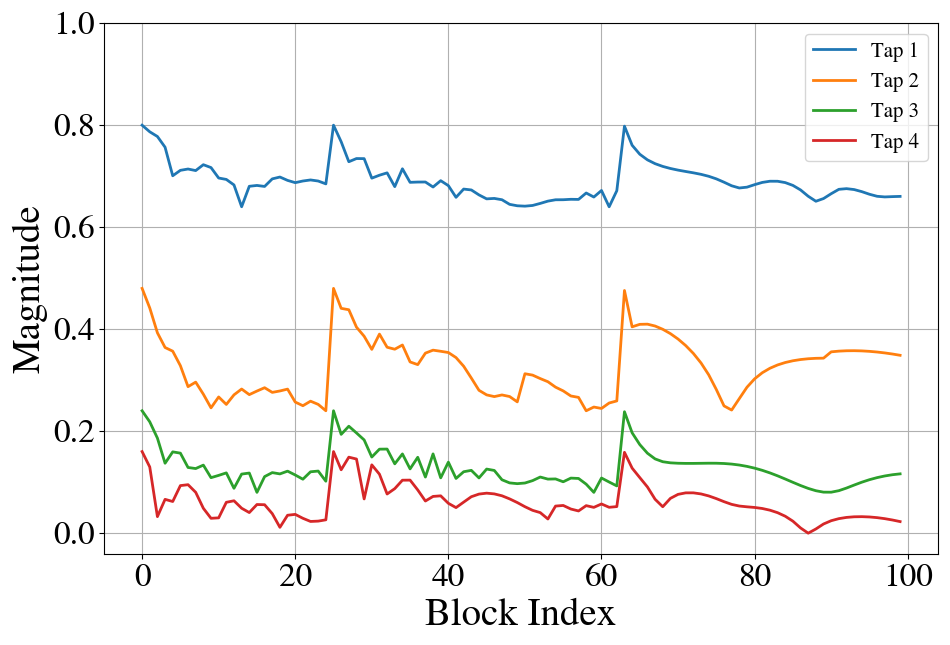}\label{fig:siso_channel}}\\
  \subfigure[MIMO Single-User Variations]
    {\includegraphics[width=0.8\linewidth]{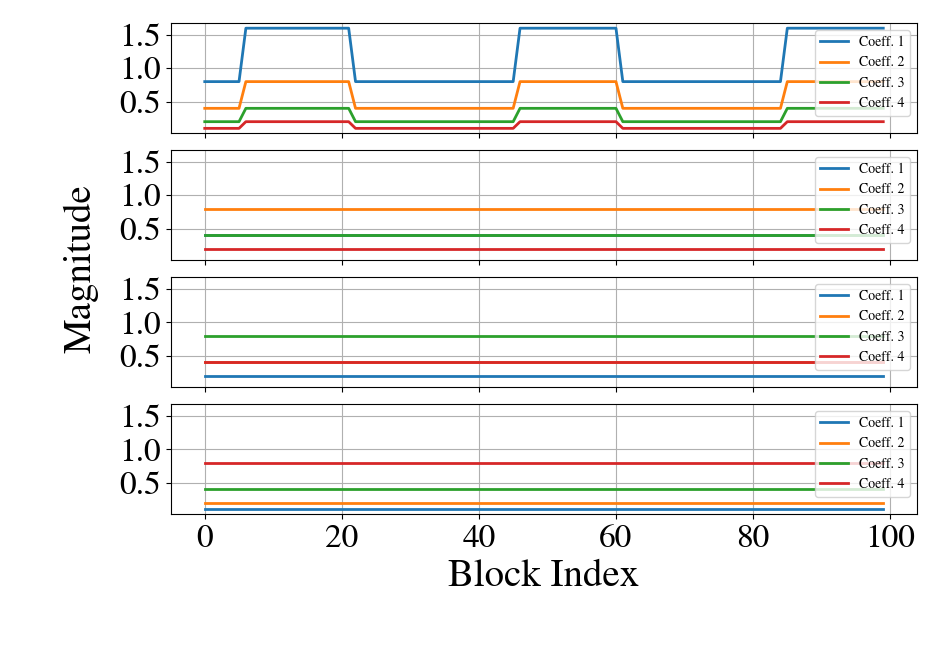}\label{fig:Mimo distorted channel}}\\
  \subfigure[MIMO Multi-User Variations]
  {\includegraphics[width=0.8\linewidth]{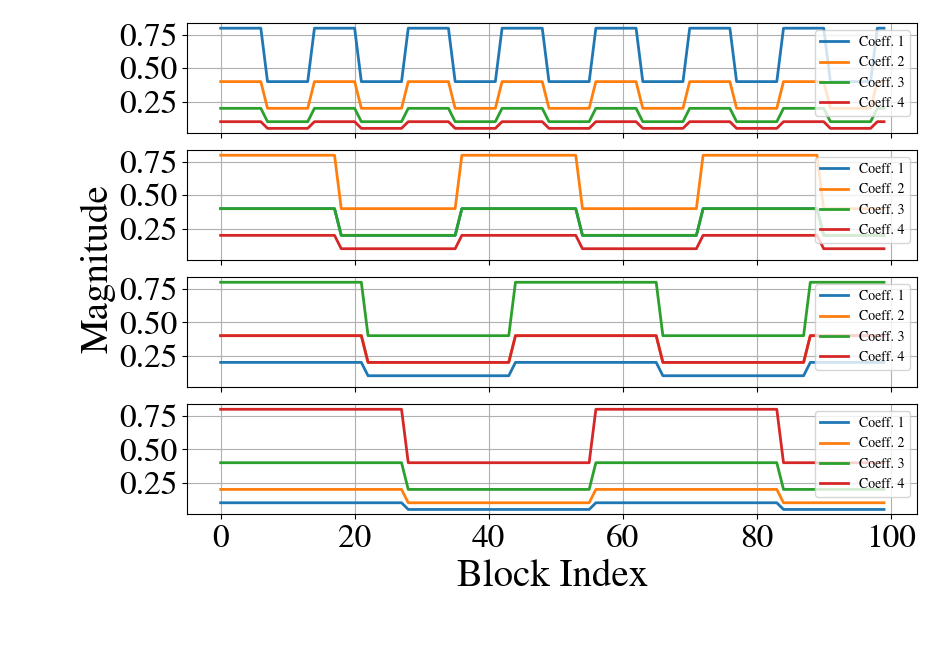}\label{fig:mimo_channel}}\\
    \subfigure[MIMO COST2100]
    {\includegraphics[width=0.8\linewidth]{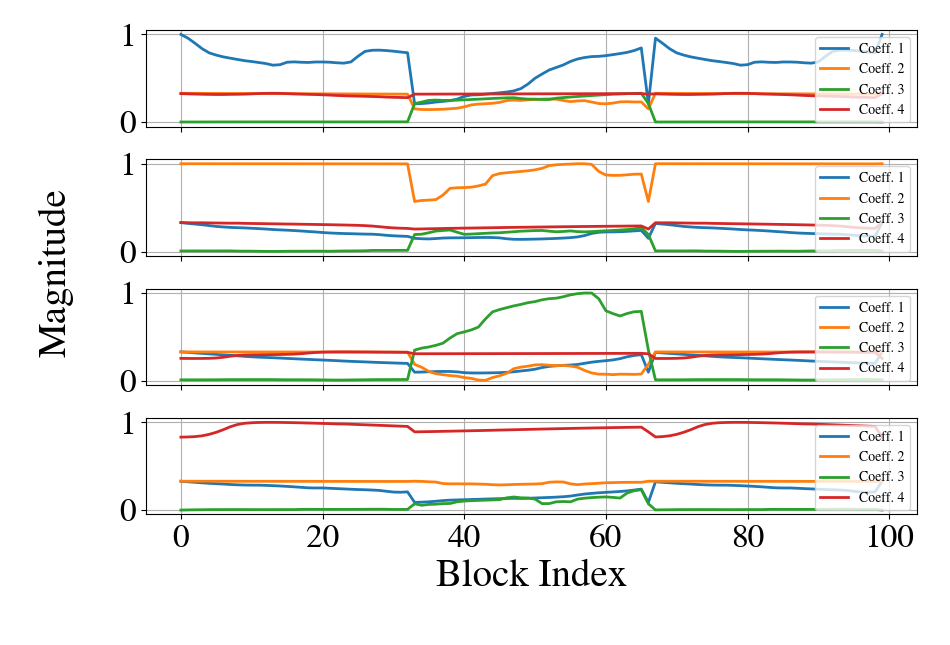}\label{fig:Mimo Cost2100 model channel}}
    \caption{Channel variation profiles. }
    \label{fig:AllChannels}
\end{figure}

Our first numerical study considers a simple \ac{siso} case, with the purpose of depicting the potential of our suggested scheme in maintaining accurate detection of deep receivers in time-varying channels while reducing the computational overhead. 
We employ a finite memory channel with four taps and \ac{awgn}. We transmit symbols from the \ac{bpsk} constellation only. The temporal variations, illustrated in Fig.~\ref{fig:siso_channel}, are simulated based on the COST2100~\cite{liu2012cost} channel using the $5$ GHz indoor hall setting. We transmit $T=100$ blocks  of $\Blklen^{\rm tran}= 10^4$ symbols each. 

We test two DNN-based receivers for the \ac{siso} setting:
\begin{itemize}
    \item ViterbiNet proposed in \cite{shlezinger2019viterbinet}, which augments the Viterbi algorithm with learnable parameters, to account for inaccurate or unknown channel knowledge. 
    \item A \ac{rnn} compromised of a sliding-window \acl{lstm} layer with a fully connected output layer. 
\end{itemize} 
For the parameters-compact ViterbiNet, which can converge using a small dataset of only a few hundred symbols, we consider $\Blklen^{\rm pilot}=500$ pilots followed by $\Blklen^{\rm info}= 9,500$ information symbols. The \ac{rnn}, on the hand, is more data consuming, thus we use $\Blklen^{\rm pilot}=1,000$ and $\Blklen^{\rm info}= 9,000$.

\subsubsection{Memoryless MIMO Channels}
The main bulk of our numerical study considers an uplink \ac{mimo} setting comprised of four users communicating with a receiver equipped with four  antennas using \ac{bpsk} signals. This scenario allows us to compare the considered drift detection mechanisms, as well as unstructured and modular asynchronous online learning. Transmission spans $T=100$ blocks arising from  three different forms of temporal variations: 
$(i)$ synthetic single-user variations (Fig.~\ref{fig:Mimo distorted channel}), representing, e.g., a single mobile user; 
$(ii)$ synthetic multi-user variations (Fig.~\ref{fig:mimo_channel}), where all four users exhibit bursty channel variations;
$(iii)$ COST2100 channel~\cite{liu2012cost} using the 5 GHz indoor hall setting  (Fig.~\ref{fig:Mimo Cost2100 model channel}) for each per user channel.

We use two deep receivers for the \ac{mimo} setting:
\begin{itemize}
    \item DeepSIC  \cite{shlezinger2019deepsic} is a modular \ac{dnn} architecture based on soft interference cancellation receiver. DeepSIC uses \ac{dnn} blocks to output the symbol estimation for each user at each iteration, thus is suitable for modular re-training. 
    \item A fully connected \ac{dnn} comprised of a multi-layered perceptron with a ReLU activation in its hidden layer, and an output  layer with a softmax activation. 
\end{itemize}
For all \ac{mimo} channels, the number of pilot symbols in each block is $\Blklen^{\rm pilot}=2000$.

\subsubsection{Training Methods}
We compare the performance achieved with the asynchronous online learning to two methods of synchronous training: 
\begin{itemize}
    \item {\em Always} - conventional synchronous online training, that re-trains on each consecutive block. The performance of this computationally-intensive approach serves as a lower bound on the achievable error-rate.
    \item {\em Periodic} -  synchronous online training that trains once every $10$ blocks, instead of every block. It benefits from reduced complexity but suffers from  the inability to choose when to issue re-training as in our suggested asynchronous  framework.
\end{itemize}

The asynchronous methods re-train in the consecutive block following a drift detection. To ensure fair comparison, we impose a constraint on the asynchronous online learning method via the setting of $C$ in \eqref{eqn:Problem}, stipulating that the number of re-training iterations does not surpass a predetermined value (ensuring fair comparison with periodic synchronous online learning). 
To report the number of re-training operations carried out, we write the average number of re-training operations over a horizon of $T$ blocks with square brackets. For instance, ${\rm DDM[9.2]}$ implies that DDM used $9.2$ re-training operations on average over the entire transmission.
 
\subsection{SISO Channel Results}
\label{ssec:SimSISO}
Our first numerical study aims at showcasing the potential of asynchronous online learning in approaching the performance of synchronous online learning while reducing the number of re-trainings. We adopt the \ac{siso} settings in Subsection~\ref{ssec:SimSetup}, and employ only the mechanisms from traditional \ac{ml} literature, i.e., the \ac{ddm} (Algorithm~\ref{alg:ddm}) and the \ac{pht} (Algorithm~\ref{alg:pht}).

\begin{figure}
\begin{center}
      \subfigure[ViterbiNet]
    { 
    \includegraphics[width=0.85\linewidth]{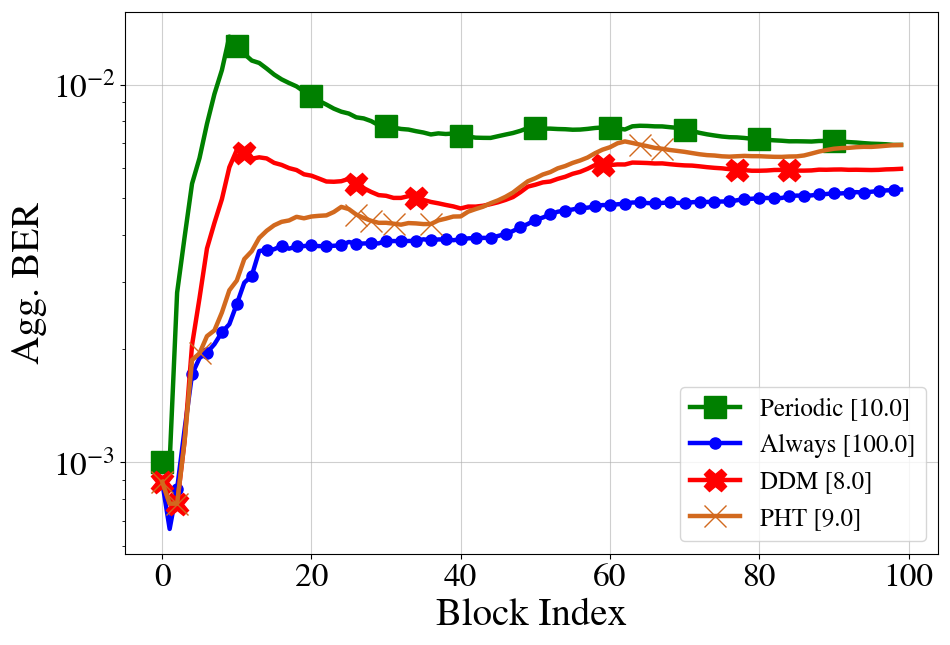}\label{fig:viterbinet}}
  \subfigure[RNN]
  { 
  \includegraphics[width=0.85\linewidth]{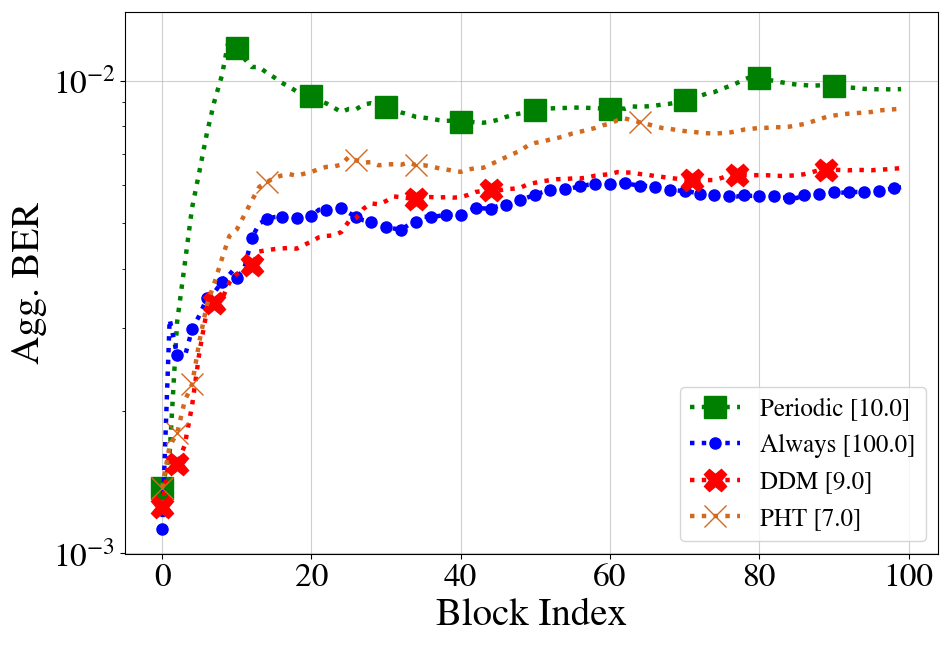}}\label{fig:rnn}
  \caption{Aggregated \ac{ber} versus block. SISO COST2100 Channel. Markers indicate re-training blocks. }
  \label{fig:SISO_BerVsBlock}
\end{center} 
  \end{figure}

We estimate the cumulative probability $\sum_{i= \Blklen^{\rm pilot} + 1}^{\Blklen^{\rm tran}} \Pr\left( \hat{\myVec{s}}_i[t] \neq  \myVec{s}_i^{\rm info}[t] \right)$ from \eqref{eqn:Problem} by calculating the accumulated  \ac{ber} as a function of the block index at an \ac{snr} of $12$ dB. The resulting \ac{ber} evolution for all considered adaptation schemes for either  ViterbiNet or the \ac{rnn} architectures are reported in Fig.~\ref{fig:SISO_BerVsBlock}. For both architectures, the gains of applying asynchronous learning is apparent. Most notably, adapting based on drifts detected in the receiver via \ac{ddm} (instead of based on the channel distribution via \ac{pht}) yields performance that closely matches synchronous online learning, while reducing the number of re-training steps by a factor of more than $\times10$ (from $100$ online training operations to only $9$).  Compared to periodic re-training, the schemes have similar complexity but allow one to select when to re-train, achieving overall lower \ac{ber}. Comparing these figures to the temporal variations depicted in Fig.~\ref{fig:siso_channel}, we observe that the detected drifts (marked with markers for each method) approximately correspond to the abrupt variations in the channel (e.g., blocks $t=25$ and $t=65$).

\begin{figure}
  \centering 
    \subfigure[ViterbiNet]
  {\includegraphics[width=0.85\linewidth]{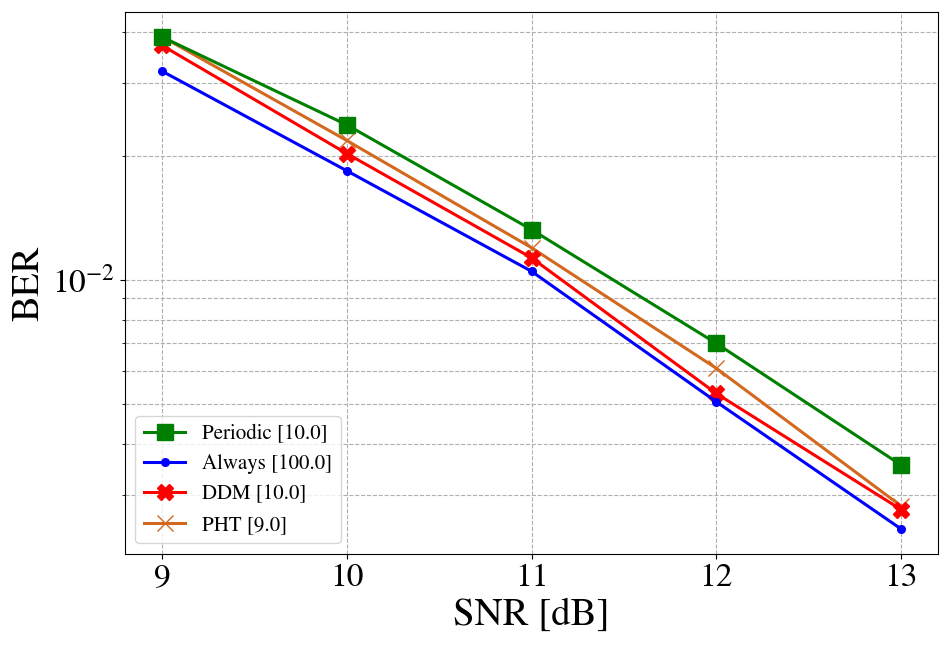}\label{fig:ViterbiNet SNR}}
    \subfigure[RNN]
  {\includegraphics[width=0.85\linewidth]{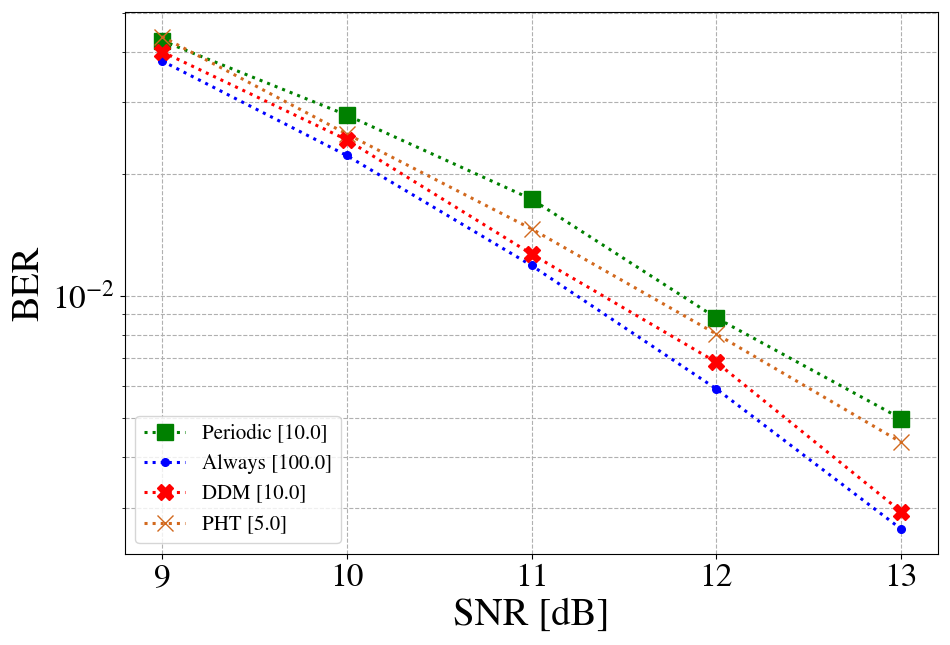}\label{fig:RNN SNR}}
  \caption{Average \ac{ber} versus SNR. SISO COST2100 Channel. }
  \label{fig:siso_performance_snrs}
  \end{figure}

Next, we repeat the above experiment for multiple \acp{snr} in the range of $9-13$dB, while using the same channel profile, and measure the average \ac{ber} after $T=100$ blocks. Fig.~\ref{fig:siso_performance_snrs} presents the  results. Note that for ViterbiNet, the \ac{ddm} mechanism allows one to reduce the complexity overhead as compared to synchronous online training, while also improving \ac{ber} as compared to the periodic training. However, the \ac{pht} shows little improvement as compared to the periodic training. These observations are not consistent, as for the \ac{rnn} architecture, both mechanisms improve on the fallacies of the synchronous and periodic online training, offering a robust trade-off over all considered noise levels. These results indicate that asynchronous online learning is able to facilitate the adaptation of deep receivers under time-varying conditions, and illustrate that detection mechanisms have a different impact for each architecture.

\subsection{Multi-User \ac{mimo} Channels}
\label{ssec:SimMIMO}
We proceed to comparing the different asynchronous online learning methods proposed in Section~\ref{sec:soft_modular_drift_detection}, as well as measuring the gains of modular asynchronous learning detailed in Section~\ref{sec:modular}. To that aim, we consider the \ac{mimo} settings from Subsection~\ref{ssec:SimSetup}. Unlike the previous study, here we not only use conventional drift detection mechanisms, but also our proposed soft-output detector based on the Hotelling test (HT).
 
\begin{figure}
  \centering 
      \subfigure[DeepSIC (Unstructured)]
    { 
    \includegraphics[width=0.85\linewidth]{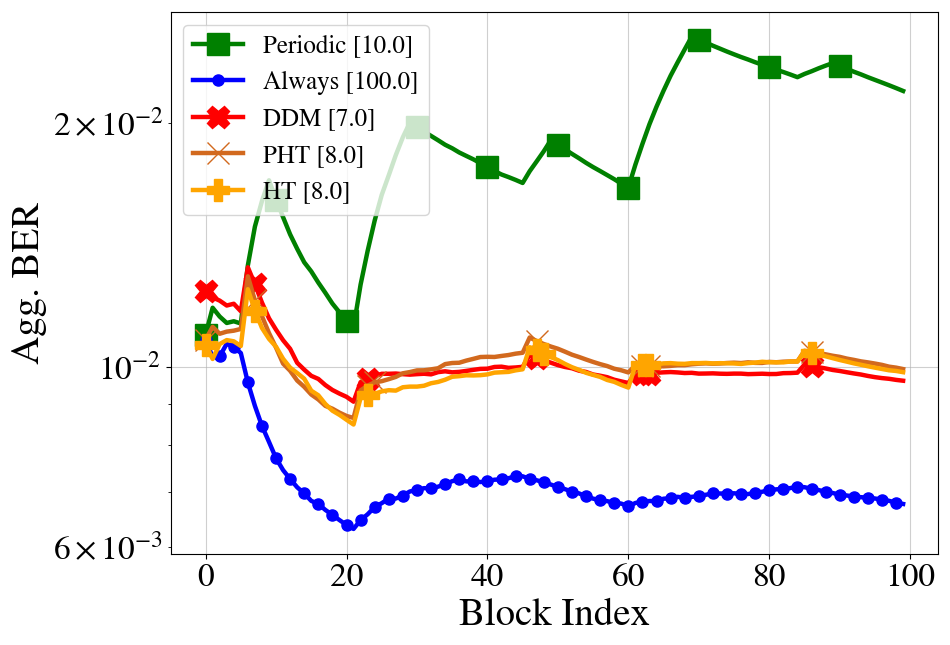}\label{DeepSIC whole model retraining synthetic}}
  \subfigure[Fully connectied \ac{dnn} (Unstructured)]
  { 
  \includegraphics[width=0.85\linewidth]{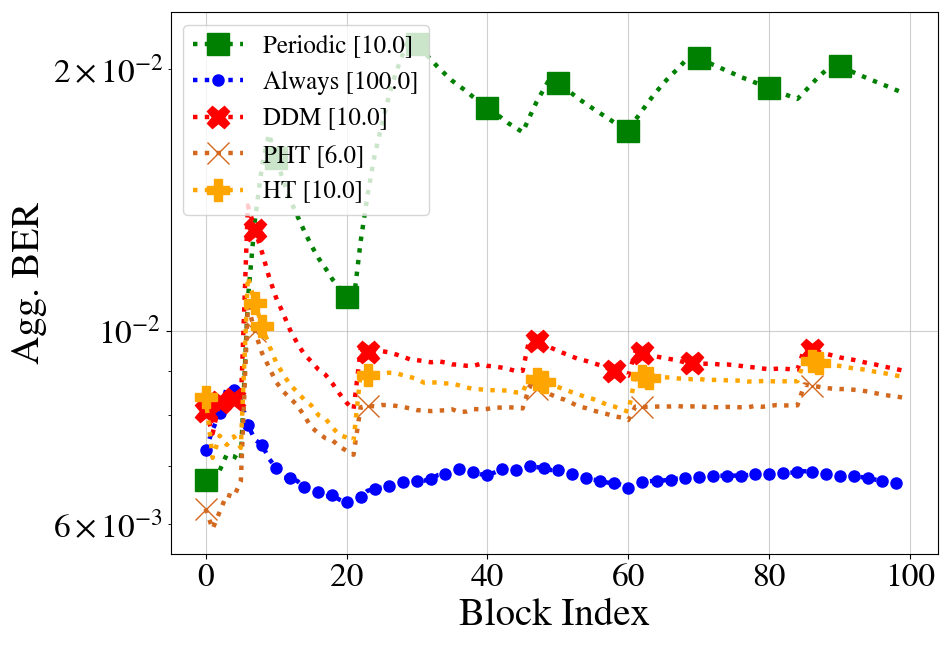}\label{DNN model retraining distorted}} 
      \subfigure[DeepSIC (Modular)]
    { 
    \includegraphics[width=0.85\linewidth]{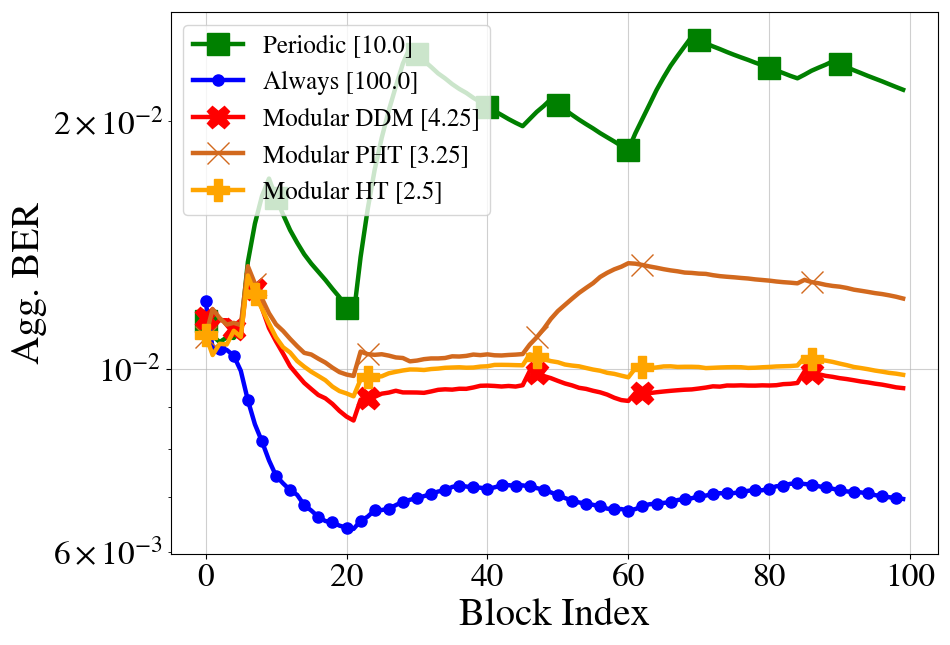}\label{DeepSIC modular retraining synthetic}}
  \caption{Aggregated \ac{ber} versus block. MIMO Single-User Variations Channel. Markers indicate re-training blocks.}
  \label{fig:MIMOSU_BerVsBlock}
  \end{figure}

\begin{figure*}
  \centering
    \subfigure[DeepSIC (Unstructured)]
  {\includegraphics[width=0.3 \linewidth]{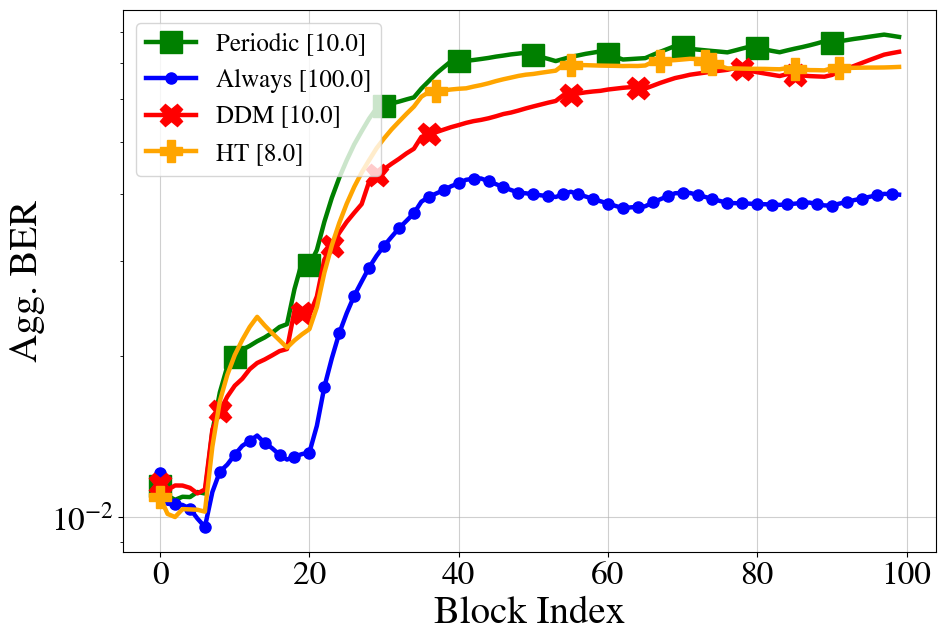}\label{fig:mimo_multi_distorted_deepsic_model}}
    \subfigure[Fully connected DNN]
  {\includegraphics[width=0.3 \linewidth]{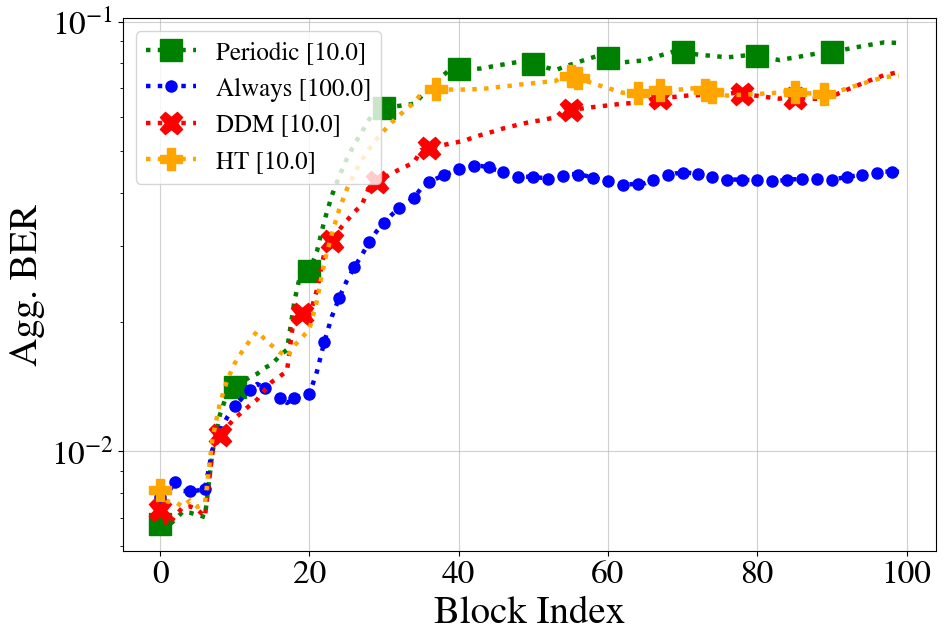}\label{fig:mimo_multi_distorted_dnn}}
  \subfigure[DeepSIC (Modular)]
  {\includegraphics[width=0.3 \linewidth]{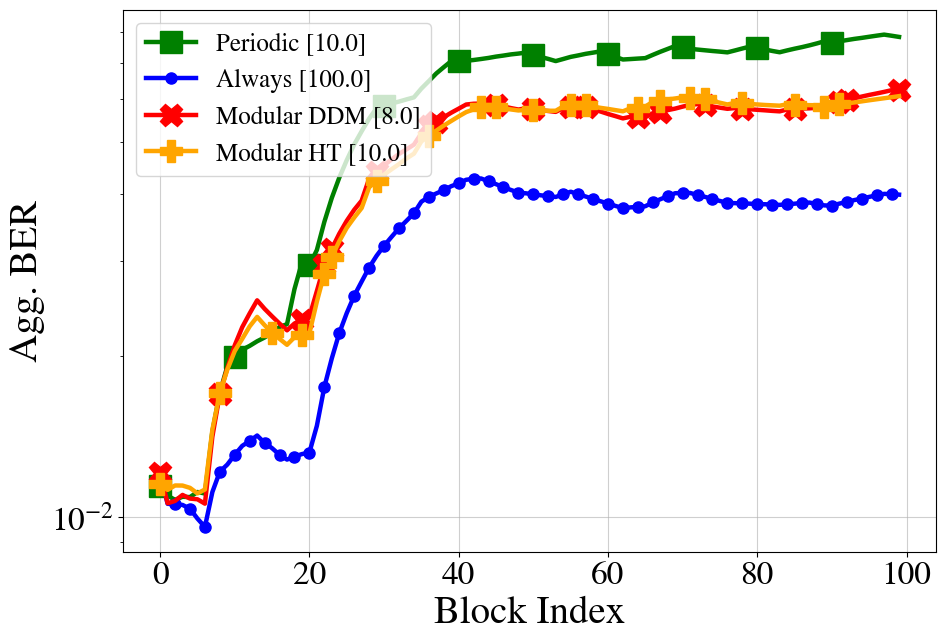}\label{fig:mimo_multi_distorted_deepsic_modular}}
  \caption{Aggregated \ac{ber} versus block. MIMO Multi-User Variations Channel. Markers indicate re-training blocks.}
  \label{fig:channels}
\end{figure*}

\subsubsection{Synthetic Single-User Variations}
We first consider a channel in which the variations are exhibited by a single user only (Fig.~\ref{fig:Mimo distorted channel}). 
We report in Fig.~\ref{fig:MIMOSU_BerVsBlock} the aggregated \ac{ber} versus the block index under \ac{snr} of $12$ dB for two deep receiver architectures, DeepSIC (Fig.~\ref{DeepSIC whole model retraining synthetic}) and the fully-connected \ac{dnn} (Fig.~\ref{DNN model retraining distorted}). We observe that all drift detection methods yield a substantial decrease in \ac{ber} with fewer retraining iterations compared to the periodic approach. The fact that all drift detectors yield relatively similar performance in unstructured asynchronous online learning stems from the infrequent and abrupt variations, yielding notable drifts in all mechanisms.

Since DeepSIC is a modular deep receiver architecture, one would expect that under single-user variations only some of the underlying modules of it should be adapted. The corresponding Fig.~\ref{DeepSIC modular retraining synthetic} depicts the \ac{ber} results for the modular asynchronous algorithm suggested in Subsection~\ref{sec:modular}. It appears that both \ac{ddm} and HT methods obtained the same \ac{ber} as in the unstructured case (Fig.~\ref{DeepSIC whole model retraining synthetic}), while further reducing the number of re-training operations by adapting only parts of the deep receiver. Notably, while \ac{ddm} outperforms HT in terms of average \ac{ber}, HT demonstrates superior efficiency, reducing the number of training rounds by a factor of $\times1.3$ as compared to \ac{ddm} and a factor of $\times40$ as compared to the synchronous {\em always} baseline. An exception to this trend was observed with \ac{pht}, which exhibits inferior performance relative to unstructured online learning, indicating that detecting drifts by evaluating the statistics of internal features is less reliable in identifying when re-training is needed compared to observing the deep receiver itself.


\subsubsection{Synthetic Multi-User Variations}
Next, we  consider a more dynamic scenario of multiple moving users, as in Fig.~\ref{fig:Mimo distorted channel}. Now the channel variations are still associated with specific modules in DeepSIC, however the relations between them is more complex. We again consider an \ac{snr} of $12$ dB, reporting the resulting aggregated \ac{ber} versus block achieved using DeepSIC trained online in an unstructured manner and in modular manner, and the fully-connected unstructured \ac{dnn},  in Fig.~\ref{fig:channels}. 

In line with previous results, a \ac{ber} decrease is observed in Fig.~\ref{fig:channels} when employing the asynchronous drift methods that utilize the deep receiver outputs, both in DeepSIC and the fully-connected \ac{dnn}. The unsupervised \ac{pht} fails in achieving reliable performance within the limited amount of allowed re-training operations as the distribution constantly changes, and therefore its performance is omitted from the figures. Here the re-training number is similar for the drift detection mechanisms and the periodic training scheme, but a reduction is observed in all methods. The gains are most significant in the modular case depicted in Fig. ~\ref{fig:mimo_multi_distorted_deepsic_modular}. This indicates that modular asynchronous online learning is beneficial even when the channel variations cannot be consistently attributed to a single sub-module, as in the study reported in Fig.~\ref{DeepSIC modular retraining synthetic}.

\subsubsection{COST2100 Channel}
We conclude our numerical evaluation by considering physically-compliant \ac{mimo} channels obtained using COST2100. The channel variations in Fig.~\ref{fig:Mimo Cost2100 model channel} corresponds to different users walking in an indoor environment, and include both smooth drifts as well as occasional bursty variations that necessitate re-training. 

\begin{figure}
  \centering 
      \subfigure[DeepSIC (Unstructured)]
    { 
    \includegraphics[width=0.85\linewidth]{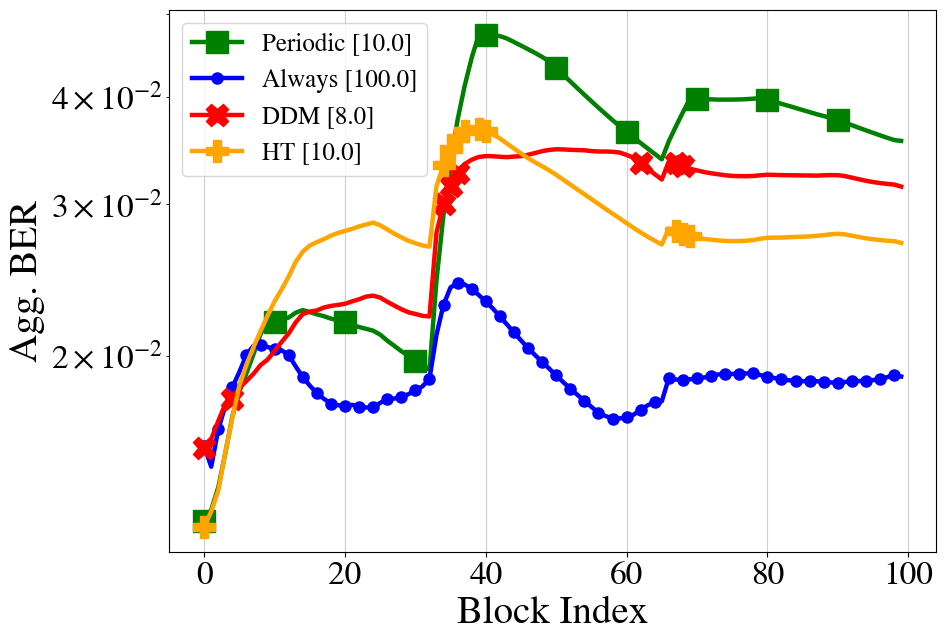}\label{DeepSIC whole model retraining cost}} 
  \subfigure[Fully connected \ac{dnn} (Unstructured)]
  { 
  \includegraphics[width=0.85\linewidth]{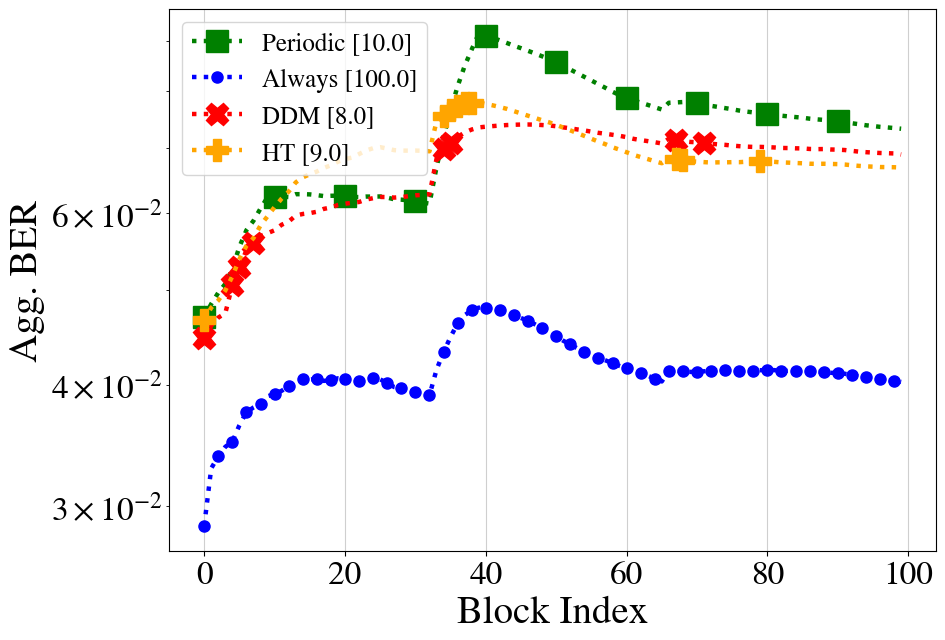}\label{DNN whole model retraining cost}}
      \subfigure[DeepSIC (Modular)]
    { 
    \includegraphics[width=0.85\linewidth]{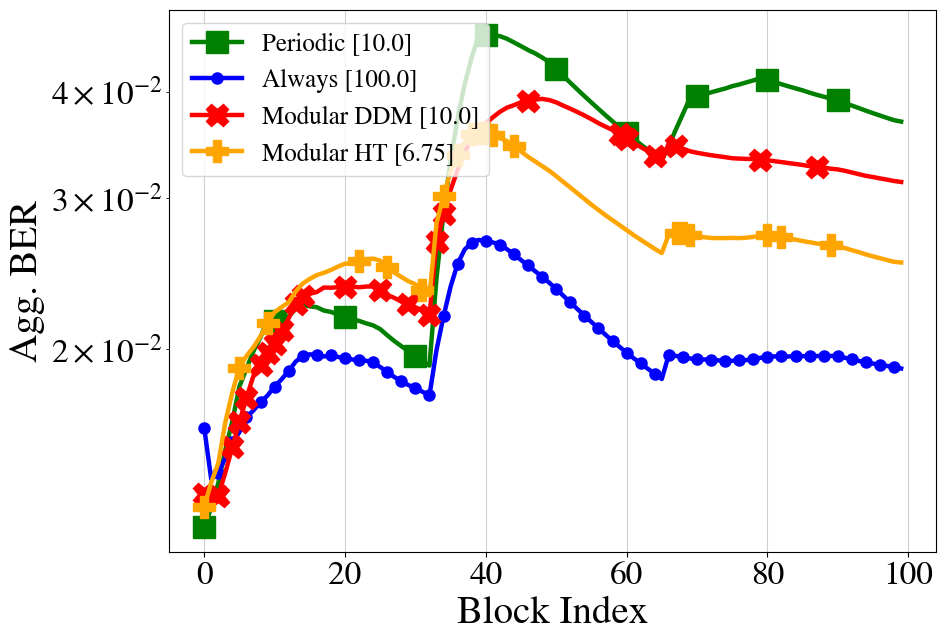}\label{DeepSIC modular retraining cost}}
  \caption{Aggregated \ac{ber} versus block. MIMO COST2100 Channel. Markers indicate re-training blocks.}
  \label{fig:MIMOCCOST_BerVsBlock}
  \end{figure}

 The aggregated \acp{ber} for unstructured asynchronous online learning of both deep receivers, as well as for modular online learning of DeepSIC, are reported in  Figs.~\ref{DeepSIC whole model retraining cost}-\ref{DeepSIC modular retraining cost}, respectively. Similarly to previous results, we see a deterioration in the \ac{ber} performance for all considered architectures as compared to synchronous online training while reducing the operations by a factor of $\times 10$. Here, our proposed HT achieves the best performance among all drift detection mechanisms, for both unstructured and modular cases. Our suggested HT-based method shines in the modular case, where it further reduces the amount of re-training of the \ac{ddm} approach, allowing a robust adaptation in combined smooth-and-bursty variations. 
 
\begin{figure}
  \centering 
      \subfigure[DeepSIC (Unstructured)]
    { 
    \includegraphics[width=0.85\linewidth]{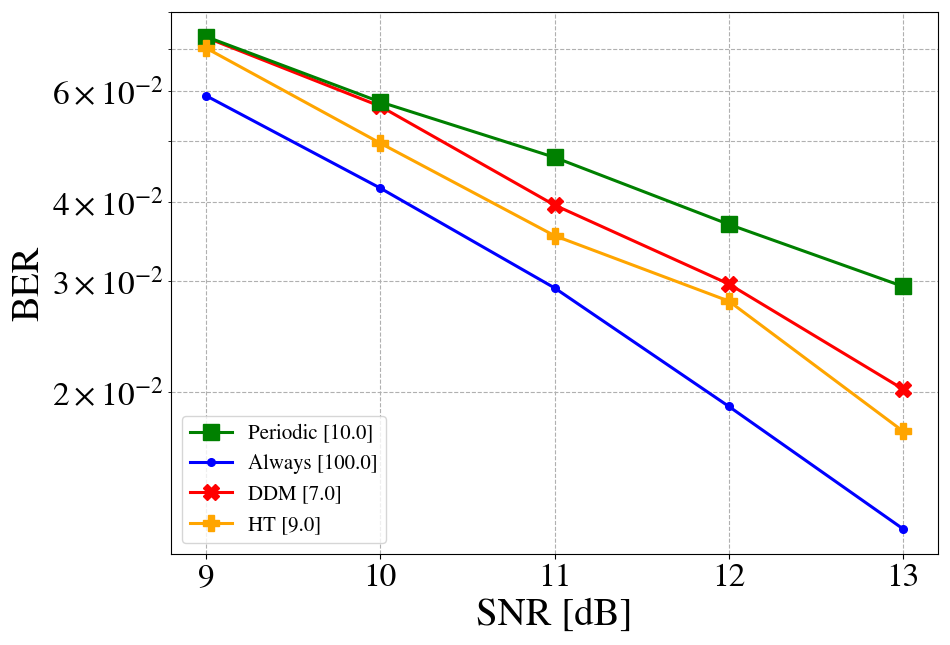}\label{DeepSIC whole model retraining cost snr}} 
  \subfigure[Fully connected \ac{dnn} (Unstructured)]
  { 
  \includegraphics[width=0.85\linewidth]{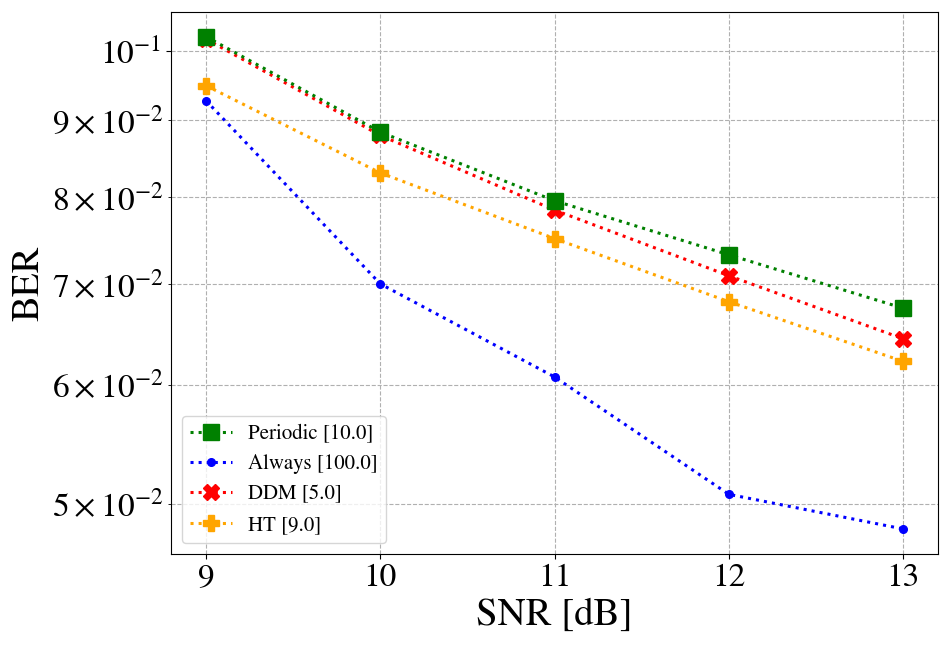}\label{DNN whole model retraining cost snr}}
      \subfigure[DeepSIC (Modular)]
    { 
    \includegraphics[width=0.85\linewidth]{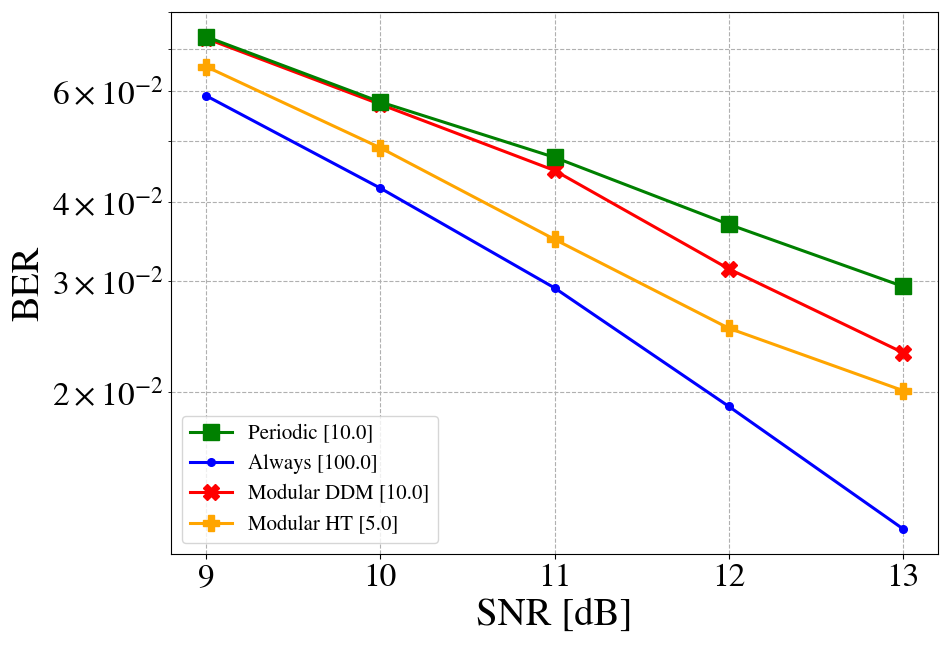}\label{DeepSIC modular retraining cost snr}}
  \caption{Average \ac{ber} versus \ac{snr}. MIMO COST2100 Channel.}
  \label{fig:MIMOCCOST_BerVsSNR}
  \end{figure}

Generalizing the above results to multiple \acp{snr}, as seen for both receivers in Fig.~\ref{fig:MIMOCCOST_BerVsSNR}, shows that the HT-based mechanism is indeed more robust for different levels of noise, and that its gains in performance and complexity reduction are consistent across all \acp{snr}. The most notable gains are again observed when adopting modular asynchronous online learning (Fig.~\ref{DeepSIC modular retraining cost snr}). Still, we note the notable reduction in computational complexity provided by asynchronous online learning leads to some performance degradation compared to re-training synchronously on each coherence interval. This performance degradation observed for DeepSIC in Figs.~\ref{DeepSIC whole model retraining cost snr} and \ref{DeepSIC modular retraining cost snr} is around  $1$ dB in \ac{snr}, while the complexity reduction is by over $11\times$ (unstructured case) and $20\times$ (modular case) less parameters re-trained.

\vspace{-0.2cm}
\section{Conclusion}
\label{sec:conclusion}
We proposed a framework for re-training deep receivers online in an asynchronous manner, building on the empirical observation that not every channel variation necessitates re-training. To enable such asynchronous online learning, we studied drift detectors, adapting two conventional mechanisms -- \ac{ddm} and \ac{pht} -- and proposing two soft-decision mechanisms. We proposed to exploit the modular structure of hybrid model-based/data-driven receivers to reduce complexity by re-training only part of the sub-modules. Our proposed asynchronous modular online learning framework  showed improved accuracy with notable computational burden reduction compared to re-training periodically, and with minimal accuracy loss compared to  online training at every coherence duration. 
	\bibliographystyle{IEEEtran}
	\bibliography{IEEEabrv,refs}

\end{document}